\documentclass[11pt]{article}

\usepackage{subfigure}
\usepackage{amssymb}
\usepackage{amsfonts}
\usepackage{bm}
\usepackage{color}
\usepackage{calc}
\usepackage[latin1]{inputenc}
\usepackage{amsfonts}
\usepackage{amsmath}
\usepackage[english]{babel}
\usepackage[T1]{fontenc}
\usepackage{verbatim}
\usepackage{subfig}
\usepackage{enumerate}
\usepackage{graphicx}
\usepackage{natbib}
\usepackage{url}
\usepackage{algorithmic,algorithm}
\usepackage{placeins}

\def\CC{{\mathbb{C}}}

\def\RR{{\mathbb{R}}}

\def\ZZ{{\mathbb{Z}}}

\def\G{{\mathcal{G}}}
\def\H{{\mathcal{H}}}
\def\I{{\mathcal{I}}}

\def\L{{\mathcal{L}}}
\def\O{{\mathcal{O}}}

\def\x{{\bm x}}

\def\x{{\mathbf x}}

\def\v{{\mathbf v}}
\def\w{{\mathbf w}}
\def\V{{\mathbf V}}
\def\0{{\mathbf 0}}

\def\bA{{\bf A}}

\def\bG{{\bf G}}

\def\bP{{\bf P}}

\def\bI{{\bf I}}

\def\bK{{\bf K}}

\def\hzeta{\widehat{\zeta}}

\def\Dpartial#1#2{ {\partial #1 \over \partial #2} }

\def\Bmp#1{ \begin{minipage}{#1} }
\def\Emp{ \end{minipage} }
\def\Bmpc#1{ \begin{minipage}[c]{#1} }
\def\Bmpt#1{ \begin{minipage}[t]{#1} }
\def\Bmpb#1{ \begin{minipage}[b]{#1} }

\DeclareMathOperator{\diag}{diag} 

\newcommand{\ci}{\mathrm{i}}
\newcommand{\dd}{\mathrm{d}}
\newcommand{\ee}{\mathrm{e}}

\renewcommand{\O}{O} 
\newcommand{\pvint}{\mbox{p.v.}\!\!\int}
\newcommand{\Gt}{{\widehat{\Gamma}}}

\begin{document}
\title{Finite rotating and translating vortex sheets}

\author{Bartosz Protas$^{1, }$\thanks{Email address for correspondence: bprotas@mcmaster.ca},  Stefan G. Llewellyn Smith$^{2,3}$ and Takashi Sakajo$^{4}$
\\ \\ 
$^1$ Department of Mathematics and Statistics, McMaster University \\
Hamilton, Ontario, L8S 4K1, Canada
\\ \\
$^2$ Department of Mechanical and Aerospace Engineering, \\ 
Jacobs School of Engineering, UCSD, \\ La Jolla CA 92093-0411, USA
\\ \\ 
$^3$ Scripps Institution of Oceanography, UCSD, \\ La Jolla CA 92093-0213, USA
\\ \\ 
$^4$ Department of Mathematics, Kyoto University \\ 
Kitashirakawa Oiwake-cho, Sakyo-ku, Kyoto, 606-8502, Japan
}

\date{\today}

\maketitle

\begin{abstract}
  We consider the rotating and translating equilibria of open finite
  vortex sheets with endpoints in two-dimensional potential flows. New
  results are obtained concerning the stability of these equilibrium
  configurations which complement analogous results known for
  unbounded, periodic and circular vortex sheets. First, we show that
  the rotating and translating equilibria of finite vortex sheets are
  linearly unstable. However, while in the first case unstable
  perturbations grow exponentially fast in time, the growth of such
  perturbations in the second case is algebraic. In both cases the
  growth rates are increasing functions of the wavenumbers of the
  perturbations.  Remarkably, these stability results are obtained
  entirely with analytical computations. {Second,} we obtain and
  analyze equations describing the time evolution of a straight vortex
  sheet in linear external fields.  {Third,} it is demonstrated
  that the results concerning the linear stability analysis of the
  rotating sheet are consistent with the {infinite-aspect-ratio} limit
  of the stability results known for Kirchhoff's ellipse
  \citep{l93,mr08} and that the solutions we obtained accounting for
  the presence of external fields are also consistent with the
  {infinite-aspect-ratio} limits of the analogous solutions known for
  vortex patches.
\end{abstract}


\section{Introduction}

Vortex sheets are often used as idealized inviscid models of complex
vortex-dominated flows, especially those arising in the presence of
separating shear layers. While some attempts have been made to model
three-dimensional vortex sheets \citep{blp98,vsheet:Sakajo3D}, most
work has focused on two-dimensional (2D) flows that can be described
more simply. Vortex sheets have been used in classical aerodynamics
\citep{MilneThomson1973book} {and} to model fluid-structure
interactions in separated flows such as flutter
\citep{Jones2003,JonesShelley2005,Alben2009,Alben2015}. The classical
problem of sheet roll-up is also receiving renewed attention
\citep{EllingGnann2019,PullinSader2021}.

Vortex equilibria have always played a distinguished role in the study
of vortex-dominated flows, as they represent long-lived flow
structures. A lot is known about vortex sheet equilibria in idealized
setting with infinite, periodic or circular sheets
\citep{saffman-1992,MarchioroPulvirenti1993}.  On the other hand, our
understanding of key properties of equilibria involving finite open
sheets (with endpoints) is far less complete.  {The goal of this
  study is thus to fill this gap partially} by establishing a number
of new facts about two equilibria of finite open vortex sheets.

We consider the inviscid evolution of a finite vortex sheet $L(t)$.
In addition to the position $\x(t,\xi) = (x(t,\xi),y(t,\xi)) \in
L(t)$, where $\xi$ is a parameter, the circulation density of the
sheet, $\gamma(t,s(\xi))$ where $s$ is the arclength coordinate, is
also needed to describe the time evolution of the vortex sheet.  This
quantity represents the jump in the tangential velocity component
across the sheet as a function of position. In the most general case,
assuming an arbitrary parameterization $\xi$ of the sheet, the
evolution of the sheet is described by the system
\citep{LopesFilho2007}
\begin{subequations}
\label{eq:xsig}
\begin{align}
\Dpartial{\x(t,\xi)}{t} + a(t,\xi) \Dpartial{\x(t,\xi)}{\xi} & = \V(\x(t,\xi)) 
:= \pvint \bK\left(\x(t,\xi) - \x(t,\xi')\right) \sigma(t,\xi') \,\dd\xi', \label{eq:x} \\
\Dpartial{\sigma(t,\xi)}{t} + \Dpartial{ \left[a(t,\xi)  \sigma(t,\xi)\right]}{\xi}  & = 0, \label{eq:sig}
\end{align}
\end{subequations}
where the Biot-Savart kernel is defined as $\bK(\x) :=
\x^{\perp}/(2\pi \vert \x \vert)$ with $(x,y)^{\perp} := (-y,x)$ and the
symbol ``$\mbox{p.v.}$'' means that integration is understood
in Cauchy's principal-value sense. The conserved quantity is defined
as $\sigma(t,\xi) := \gamma(t,s(t,\xi)) (\dd s/\dd\xi)$, while
$a(t,\xi)$ is determined by the parameterization.  Specific forms of
parameterization which have been considered in the literature include
parameterization in terms of the arclength $s$
\citep{DevoriaMohseni2018} and in terms of the graph of a function
with $\x = [x, y(x)]^T$ \citep{MarchioroPulvirenti1993}. However, for
the particular parameterization in terms of the circulation
\begin{equation}
\Gamma(s) := \int_0^s \gamma(s')\, \dd s'
\label{eq:Gamma}
\end{equation}
contained between the sheet endpoint and the point with the arclength
coordinate $s$, we have $\sigma = \gamma (\dd s/\dd\Gamma) \equiv 1$ and $a
\equiv 0$. Then equation \eqref{eq:sig} is satisfied trivially,
whereas equation \eqref{eq:x} rewritten using the complex
representation in terms of $z = z(t,\Gamma) = x(t,\Gamma) + \ci y(t,\Gamma)$ becomes the celebrated
Birkhoff--Rott equation \citep{saffman-1992}
\begin{equation}
\Dpartial{\overline{z}(t,\Gamma)}{t} = 
\frac{1}{2\pi\ci} \pvint_{L(t)} \frac{\dd\Gamma'}{z(t,\Gamma) - z(t,\Gamma')},
\label{eq:BR} 
\end{equation}
where overbar denotes complex conjugation. The system
  \eqref{eq:xsig}, or equivalently equation \eqref{eq:BR}, represents
  a free-boundary problem in which the time-dependent shape of the
  interface (sheet) needs to be found as a part of the solution to the
  problem.

We remark that formulations employing different parameterizations of
the sheet have the same normal velocity component in \eqref{eq:x}, but
different tangential components determined by the parameterization
\citep{LopesFilho2007}. In particular, in the Lagrangian
parameterization in terms of the circulation $\Gamma$, the point
$\x(t,\Gamma)$ moves with the velocity $\V(\x(t,\Gamma))$ also in the
tangential direction. Thus, $a \neq 0$ is a measure of the departure
from Lagrangian motion. Another remarkable feature of the Lagrangian
parameterization is that the Birkhoff--Rott equation \eqref{eq:BR}
also contains information about the evolution of the circulation
density $\gamma(t,s)$ which is implicit in the definition of the
independent variable in \eqref{eq:Gamma}. 

The Birkhoff-Rott equation \eqref{eq:BR} is known to be ill-posed and
to lead to singularities in finite time \citep{vsheet:Moore}. For
these and other related reasons, this equation has been at the centre
of a lot of mathematical research and many important results are
summarized in the collection edited by \citet{c89a} and in the
monograph by \citet{mb02}.  Because of its compact form, the
Birkhoff-Rott equation \eqref{eq:BR} has been used in many numerical
studies of the evolution of vortex sheets typically involving some
form of regularization
\citep{k86a,k86b,vsheet:KrNi02,vsheet:SaOk96,DevoriaMohseni2018}.
Similarly, we will also use it, albeit without any regularization, as
the point of departure for the different analyses in the present
study. The interesting question of how to recover the circulation
density $\gamma(t,s)$ from the Lagrangian representation $z(t,\Gamma)$
will be addressed in \S\,\ref{sec:constraints}.

The velocity $\V = (u, v)$ on the right-hand side (RHS) of
\eqref{eq:x} can be expressed in complex notation  as
\begin{equation}
\forall z \in L \quad u - \ci v = \frac{1}{2\pi\ci} \pvint_{L} \frac{\gamma(s')}{z - z(s')} \,\dd s'
=  \frac{1}{2\pi\ci} \pvint_{L} \frac{\varphi(z')}{z - z'} \,\dd z',
\label{eq:V}
\end{equation}
where $\varphi(z) := \gamma(s(z)) (\dd z/\dd s)^{-1}$ is introduced so
that we can rewrite the integral as a complex integral.  It is known
that in order for the integrals in \eqref{eq:V} to be well-defined in
Cauchy's principal-value sense, the function $\varphi(z)$ must be
H\"older-continuous which also implies a similar condition on the
circulation density $\gamma(s)$ \citep{Muskhelishvili2008}.
Furthermore, in order for the velocity in \eqref{eq:V} to be bounded
everywhere on and in the neighborhood of the sheet $L$, including the
case when the point $z$ approaches the endpoints $c_1,c_2$ of the
sheet, we must have $\varphi(c_1) = \varphi(c_2) = 0$
\citep{Muskhelishvili2008}, implying that
\begin{equation}
\gamma(s(c_1)) = \gamma(s(c_2)) = 0,
\label{eq:gamma_ep}
\end{equation}
where $s(c_1)$ and $s(c_2)$ denote the arclength coordinates of the
endpoints of the sheet. Condition \eqref{eq:gamma_ep} thus defines a
class of physically-admissible circulation densities as those that
vanish at the endpoints of the sheet.

In this study we focus on two equilibria involving finite open vortex
sheets. The first is the rotating while the second is the translating
equilibrium, also known as the Prandtl-Munk vortex. While the linear
stability of the straight infinite and the closed circular sheet has
been understood for a long time
\citep{MichalkeTimme1967,saffman-1992}, little is known about the
stability properties of open finite sheets.  As the {first} main
result of the paper, we show that the rotating and translating
equilibria of finite vortex sheets are linearly unstable. However,
while in the first case unstable perturbations grow exponentially fast
in time, the growth of unstable perturbations in the second case is
algebraic. In both cases the growth rates are increasing functions of
the wavenumbers of the perturbations. Remarkably, these stability
results are obtained entirely with analytical computations. As
{the second} contribution, we also obtain and analyze equations
describing the time evolution of a straight vortex sheet in a linear
external velocity field.

\citet{batchelor-1988} argued that the rotating equilibrium of the
vortex sheet can be obtained as an {infinite-aspect-ratio} limit of
Kirchhoff's ellipse in which circulation is conserved. We demonstrate
that this analogy goes further and in fact also applies to many key
findings of the present study. More precisely, as our final
contribution, we show that the results concerning the linear stability
analysis of the rotating sheet are consistent of the {infinite-aspect-ratio}
limit of the stability results known for Kirchhoff's ellipse
\citep{l93,mr08}.

The structure of the paper is as follows. In the next section we
recall the rotating and translating equilibria of finite open vortex
sheets. Next in \S\,\ref{sec:stability} we carry out the linear
stability analysis of these equilibria. In \S\,\ref{sec:timedep} we
construct time-dependent generalizations of these equilibria in the
presence of linear strain and shear.  Finally in
\S\,\ref{sec:relation} we demonstrate that most of the results
reported in \S\,\ref{sec:stability} and \S\,\ref{sec:timedep} are
consistent with the {infinite-aspect-ratio} limits of solutions
involving rotating vortex ellipses. A discussion and conclusions are
presented in \S\,\ref{sec:final}, while some additional technical
material is given in Appendix \ref{sec:ONeil}.

\section{Two relative equilibria of a straight vortex sheet}
\label{sec:eq}

In this section we {recall some basic facts about} the rotating
and translating equilibrium configurations of a single open sheet. The
rotating equilibrium is mentioned by \citet{batchelor-1988} as a limit
of the Kirchhoff elliptical vortex, whereas the translating
equilibrium is known as the Prandtl-Munk vortex \citep{Munk1919} and
has received attention in classical aerodynamics as a simple model for
elliptically loaded wings.  Interestingly, as proved in
\citet{LopesFilho2003,LopesFilho2007}, while the rotating equilibrium
can be interpreted as a weak solution of the 2D Euler equation, the
translating equilibrium cannot. The rotating equilibrium was recently
generalized to configurations involving multiple straight segments
with one endpoint at the centre of rotation and the other at a vertex
of a regular polygon by \citet{ProtasSakajo2020}. By allowing for the
presence of point vortices in the far field
\citet{ONeil2018b,ONeil2018a} were able to find more general
equilibria involving multiple vortex sheets, including curved ones, in
both rotating and translating frames of reference.

\subsection{Rotating Equilibrium}
\label{sec:Rot_Eq}

Without loss of generality, a rotating equilibrium is sought in which
the sheet rotates anticlockwise about its centre point with angular
frequency $\Omega = 1$. The sheet can thus be described as $L(t) = L_0
e^{\ci t}$, where $L_0 := [-1,1]$, with the centre of rotation at the
origin and $L(t) = L(t+2 \pi n)$, $n \in \ZZ$. Transforming the
Birkhoff-Rott equation \eqref{eq:BR} to the rotating frame of
reference via the change of variables $Z(t,\Gamma) = z(t,\Gamma)
e^{-\ci t}$ yields
\begin{equation}
\Dpartial{\overline{Z}(t,\Gamma)}{t} = 
\frac{1}{2\pi\ci} \pvint_{L_0} \frac{\dd\Gamma'}{Z(t,\Gamma) - Z(t,\Gamma')} + \ci  \overline{Z}(t,\Gamma).
\label{eq:BRr} 
\end{equation}
Noting that the time derivative now vanishes and changing the integration
variable to $x$ (which differs from the arclength $s$ by an additive
constant) leads to
\begin{equation}
- \ci  x = 
\frac{1}{2\pi\ci} \pvint_{-1}^1 \frac{\gamma_0(x')}{x - x'} \,\dd x', \quad \forall x \in [-1,1]
\label{eq:BR0r}
\end{equation}
as a relation characterizing the rotating equilibrium. The circulation
density satisfying this equation has the form
\begin{equation}
\gamma_0(x) = 2\sqrt{1 - x^2}, \quad x \in [-1,1],
\label{eq:gamma0r}
\end{equation}
which is clearly H\"older continuous and satisfies conditions
\eqref{eq:gamma_ep}. Therefore, in this equilibrium configuration the
velocity induced by the sheet on itself (equal to $-\ci x$, which is
the opposite of the velocity due to the background rotation) is well
behaved everywhere its neighborhood. The bijective relation between
the circulation parameter $\Gamma$ and arclength $s$ (equivalently,
the coordinate $x$) for the rotating equilibrium is given by
\begin{equation}
\Gamma(x) = \int_{-1}^x \gamma_0(\xi) \, \dd\xi =  \int_{-1}^x 2\sqrt{1 - \xi^2} \,\dd\xi = 
\frac{\pi}{2} + x \sqrt{1-x^2} + \arcsin{x}.
\label{eq:Gamma0r}
\end{equation}
We note that the total circulation of the sheet is then given by $\Gt
= \Gamma(1) = \pi$. Generalizations of the equilibrium solution
described above to flows in the presence of external strain and/or
shear are described in \S\,\ref{sec:timedep}.

\subsection{Translating Equilibrium}
\label{sec:Trans_Eq}

The translating equilibrium involves a straight vortex sheet $L_0$
moving steadily in the direction perpendicular to itself with a
constant velocity $W$.  The corresponding circulation density does not
satisfy conditions \eqref{eq:gamma_ep}, so the flow velocity near the
sheet endpoints is unbounded.  The sheet in such an equilibrium
configuration can thus be described by $L(t) = L_0 - \ci t$, taking $W
= 1$.  Transforming the Birkhoff-Rott equation \eqref{eq:BR} to a
translating frame of reference via the change of variable $Z(t,\Gamma)
= z(t,\Gamma) + \ci t$ yields
\begin{equation}
\Dpartial{\overline{Z}(t,\Gamma)}{t} = 
\frac{1}{2\pi\ci} \pvint_{L_0} \frac{\dd\Gamma'}{Z(t,\Gamma) - Z(t,\Gamma')} - \ci.
\label{eq:BRt} 
\end{equation}
Then, noting that the time derivative vanishes and changing the integration
variable to $x$ we obtain
\begin{equation}
\ci =  \frac{1}{2\pi\ci} \pvint_{-1}^1 \frac{\gamma_0(x')}{x - x'}\,\dd x', \qquad \forall x \in [-1,1]
\label{eq:BR0t}
\end{equation}
as a relation characterizing the translating equilibrium. The
circulation density satisfying this equation has the form
\begin{equation}
\gamma_0(x) = \frac{2x}{\sqrt{1 - x^2}}, \qquad x \in [-1,1].
\label{eq:gamma0t}
\end{equation}
Evidently, this function is not H\"older-continuous at the
endpoints $x = \pm 1$. As a result, the velocity field induced by the
vortex sheet in such a translating equilibrium configuration is
unbounded near the endpoints where it has an inverse square-root
singularity \citep{Muskhelishvili2008}. However, as is evident from
\eqref{eq:BR0t}, on the sheet itself the induced velocity remains
bounded. The relation between the circulation parameter $\Gamma$ and
arclength $s$ (or equivalently the coordinate $x$) for the translating
equilibrium is given by
\begin{equation}
\Gamma(x) = \int_{-1}^x \gamma_0(\xi) \, \dd \xi =  \int_{-1}^x \frac{2\xi}{\sqrt{1 -\xi^2}} \,\dd\xi = - 2\sqrt{1-x^2}.
\label{eq:Gamma0t}
\end{equation}
We note that the total circulation of the sheet vanishes since
$\Gt = \Gamma(1) = 0$.

\section{Linear Stability Analysis}
\label{sec:stability}

\begin{figure}
\centering 
\includegraphics[width=0.5\textwidth]{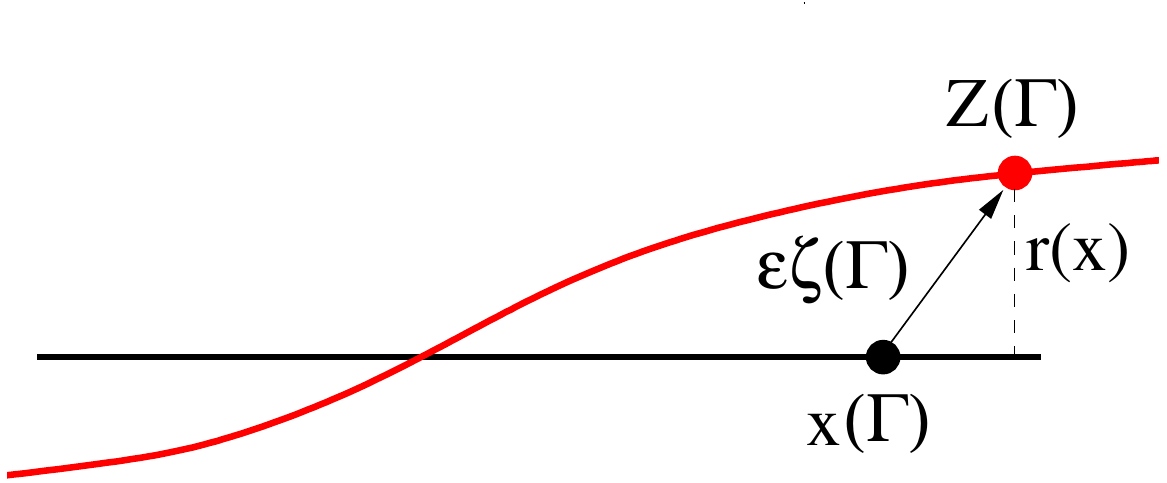}
\caption{Schematic representation of the perturbation defined by \eqref{eq:Z}.}
\label{fig:Z}
\end{figure}

In this section we analyze the stability of the equilibrium
configurations introduced in \S\,\ref{sec:Rot_Eq} and
\ref{sec:Trans_Eq} in essentially the same way in both cases. To fix
attention, we first consider the Birkhoff-Rott equation in the
rotating frame of reference \eqref{eq:BRr} and study the amplification
of infinitesimal perturbations around the equilibrium defined by
relations \eqref{eq:BR0r}--\eqref{eq:gamma0r}. We thus need to
linearize equation \eqref{eq:BRr} around this equilibrium. We write
(see Figure \ref{fig:Z})
\begin{equation}
Z(t,\Gamma) = x(\Gamma) + \epsilon \, \zeta(t,\Gamma), \quad \vert \epsilon \vert \ll 1.
\label{eq:Z}
\end{equation}
Note that while the imaginary component of the perturbation
$\zeta(t,\Gamma)$ describes the deformation of the sheet, its real
part encodes information about perturbations to the circulation
density $\gamma(s)$.

Plugging \eqref{eq:Z} into \eqref{eq:BRr}, expanding the terms
in this equation in a Taylor series in $\epsilon$ around $\epsilon =
0$ and retaining terms proportional to $\epsilon$ yields
\begin{equation}
\Dpartial{\overline{\zeta}(t,\Gamma)}{t} = \ci(\H \zeta)(t ,\Gamma) + \ci \overline{\zeta}(t,\Gamma),
\label{eq:dBRr} 
\end{equation}
where 
\begin{equation}
(\H\zeta)(t,\Gamma) :=
\frac{1}{2\pi} \mbox{f.p.} \int_0^{{\Gt}} \frac{\zeta(t,\Gamma) - \zeta(t,\Gamma')}{[x(\Gamma) - x(\Gamma')]^2}\, \,\dd\Gamma'
\label{eq:H} 
\end{equation}
is a hypersingular integral operator. {The} symbol ``$\mbox{f.p.}$''
indicates that the integral is understood in the sense of Hadamard's
finite part \citep{EstradaKanwal2012}.  In the present problem the
relation between the coordinates $\Gamma$ and $x$ in \eqref{eq:H} is
given in \eqref{eq:Gamma0r}. It is illuminating to separate equation
\eqref{eq:dBRr} into its real and imaginary parts using $\zeta =
\zeta^r + \ci \zeta^i$, leading to
\begin{subequations}
\label{eq:dBRr2}
\begin{align}
\Dpartial{\zeta^r(t,\Gamma)}{t} &= - \left(\H\zeta^i\right)(t,\Gamma) + \zeta^i(t,\Gamma), \label{eq:dBRr2r} \\
\Dpartial{\zeta^i(t,\Gamma)}{t} &= - \Big(\H\zeta^r\Big)(t,\Gamma) - \zeta^r(t,\Gamma). \label{eq:dBRr2i} 
\end{align}
\end{subequations}
The integro-differential system \eqref{eq:dBRr2} describes the
evolution of infinitesimal perturbations to the equilibrium. Assuming
that the real and imaginary parts of the perturbation depend on time
as $\zeta^r(t,\Gamma) = \ee^{\ci \lambda t} \hzeta^r(\Gamma)$ and
$\zeta^i(t,\Gamma) = \ee^{\ci \lambda t} \hzeta^i(\Gamma)$ turns
\eqref{eq:dBRr2} into the eigenvalue problem
\begin{subequations}
\label{eq:evpBRr}
\begin{align}
\ci \lambda\, \hzeta^r(\Gamma) &= - \left(\H\hzeta^i\right)(\Gamma) + \hzeta^i(\Gamma), \label{eq:evpBRr2r} \\
\ci \lambda\, \hzeta^i(\Gamma) &= - \left(\H\hzeta^r\right)(\Gamma) - \hzeta^r(\Gamma), \label{eq:evpBRr2i} 
\end{align}
\end{subequations}
where $\lambda \in \CC$ is the eigenvalue and $\hzeta^r$, $\hzeta^i$
the corresponding eigenvectors. 

Performing the same steps for the translating equilibrium described by
\eqref{eq:BR0t}--\eqref{eq:gamma0t} leads to the linearized system
\begin{subequations}
\label{eq:dBRt2}
\begin{align}
\Dpartial{\zeta^r(t,\Gamma)}{t} &= - \left(\G\zeta^i\right)(t,\Gamma), \label{eq:dBRt2r} \\
\Dpartial{\zeta^i(t,\Gamma)}{t} &= - \Big(\G\zeta^r\Big)(t,\Gamma), \label{eq:dBRt2i} 
\end{align}
\end{subequations}
where the hypersingular integral operator $\G$ is defined as in
\eqref{eq:H}, except that the relation between the coordinates
$\Gamma$ and $x$ is now given in \eqref{eq:Gamma0t}. The corresponding
eigenvalue problem then takes the form
\begin{subequations}
\label{eq:evpBRt}
\begin{align}
\ci \lambda\, \hzeta^r(\Gamma) &= - \left(\G\hzeta^i\right)(\Gamma), \label{eq:evpBRt2r} \\
\ci \lambda\, \hzeta^i(\Gamma) &= - \left(\G\hzeta^r\right)(\Gamma). \label{eq:evpBRt2i} 
\end{align}
\end{subequations}

In principle, the linearized systems \eqref{eq:dBRr2} and
\eqref{eq:dBRt2} have been obtained in a similar manner to the
corresponding system in the case of the periodic vortex sheet
\citep{saffman-1992}, except for a difference in the form of the
kernel of the integral operator \eqref{eq:H}, more specifically how
the coordinate $x$ depends on the integration variable $\Gamma'$.  and
the presence of additional terms representing the background rotation
in \eqref{eq:dBRr2}.  In our analysis of eigenvalue problems
\eqref{eq:evpBRr} and \eqref{eq:evpBRt} below it will be convenient to
switch between parameterizations in terms of $\Gamma$ and $x$, which
will be facilitated by relations \eqref{eq:Gamma0r} and
\eqref{eq:Gamma0t}.

\subsection{Constraints on Admissible Perturbations}
\label{sec:constraints}

In order for perturbations $\zeta(t,\Gamma)$ to be physically
admissible, they have to satisfy certain conditions. In the present
problem we will require them to leave the total circulation $\Gt$ of
the sheet unchanged. Moreover, in the case of the rotating equilibrium
we will also require the associated circulation densities to satisfy
conditions \eqref{eq:gamma_ep}, so that the velocity induced by the
perturbed sheet remains everywhere bounded. In the case of the
translating equilibrium, the analogous condition will require the
perturbations to leave the type of singularity in the velocity field
induced by the sheet near its endpoints unchanged. The main difficulty
is that these conditions are naturally expressed in terms of the
perturbed circulation density {$\gamma$} which enters into
\eqref{eq:Z} implicitly via the circulation parameter $\Gamma$. We
thus need to translate these two conditions into constraints on the
functions $\zeta(t,\Gamma)$.

In the rotating or translating frame of reference the perturbed sheet
can be represented as the graph of a function $[x,r(x)]$, as in Figure
\ref{fig:Z}, where the normal displacement $r(x)$ is related to the
perturbation $\zeta(\Gamma)$ via
\begin{equation}
r(x) = \Im\,[Z(\Gamma(x))] = \epsilon\, \Im\,[\zeta(\Gamma(x))]
\label{eq:r}
\end{equation}
(for brevity, the dependence on time $t$ is omitted in this
discussion).  The circulation density $\gamma(x)$ is obtained from
$\zeta(\Gamma)$ as follows 
\begin{align}
& \left.
\begin{aligned}
\gamma(s) & = \frac{\dd\Gamma}{\dd s} = \left( \frac{\dd s}{\dd\Gamma}\right)^{-1} =
\left( \Dpartial{Z}{\Gamma} \, \Dpartial{\overline{Z}}{\Gamma} \right)^{-1/2} \\
\Dpartial{{Z}}{\Gamma} & = \Dpartial{x}{\Gamma} + \epsilon \Dpartial{\zeta}{\Gamma}
= \gamma_0^{-1} + \epsilon \Dpartial{\zeta}{\Gamma} 
\end{aligned}
\right\} \quad \Longrightarrow  \nonumber \\
&
\begin{aligned}
\gamma(s) & = \left[ \left(\gamma_0^{-1} + \epsilon \Dpartial{\zeta}{\Gamma} \right)
\left(\gamma_0^{-1} + \epsilon \Dpartial{\overline{\zeta}}{\Gamma} \right) \right]^{-1/2} \\
& = \gamma_0 - \epsilon \, \gamma_0^2 \, \Re\,\left[\Dpartial{\zeta}{\Gamma} \right] + \O(\epsilon^2)
= \gamma_0\left[1 - \epsilon \Dpartial{\zeta^r}{x} \right] + \O(\epsilon^2),
\end{aligned}
\label{eq:gamma2}
\end{align}
where we used the property $\gamma_0^{-1} \partial/\partial x =
\partial/\partial\Gamma$.

The total circulation of the perturbed sheet is
\begin{align}
\int_0^{\Gt} \, d\Gamma & = \int_{-1}^1 \frac{\dd\Gamma}{\dd s}\frac{\dd s}{\dd x}\, \dd x = 
\int_{-1}^1 \left( \Dpartial{Z}{\Gamma} \, \Dpartial{\overline{Z}}{\Gamma} \right)^{-1/2}
\left( \Dpartial{Z}{x} \, \Dpartial{\overline{Z}}{x} \right)^{1/2} \,\dd x \nonumber \\
& = \int_{-1}^1 \left[ \left(\gamma_0^{-1} + \epsilon \Dpartial{\zeta}{\Gamma} \right)
\left(\gamma_0^{-1} + \epsilon \Dpartial{\overline{\zeta}}{\Gamma} \right) \right]^{-1/2}
 \left[ \left(1 + \epsilon \Dpartial{\zeta}{x} \right)
\left(1 + \epsilon \Dpartial{\overline{\zeta}}{x} \right) \right]^{1/2}\,\dd x \nonumber \\
& = \int_{-1}^1 \left[ \gamma_0 - \epsilon \gamma_0^2 \Re\,\left(\Dpartial{\zeta}{\Gamma} \right)
+  \O(\epsilon^2) \right] \left[ 1 + \epsilon \Re\,\left(\Dpartial{\zeta}{x} \right)
+  \O(\epsilon^2) \right] \,\dd x \nonumber \\
& = \int_{-1}^1 \gamma_0(x) \,\dd x + \O(\epsilon^2),
\label{eq:dG}
\end{align}
where we have used \eqref{eq:gamma2}. This relation implies that, to
leading order in $\epsilon$, the total circulation is unaffected by
perturbations of the form \eqref{eq:Z}.

Hence the leading-order term in the expression for the perturbed
circulation density is proportional to
\begin{equation}
\gamma_0(x) \Dpartial{\zeta^r}{x}(\Gamma(x)).
\label{eq:gamma1}
\end{equation}
In the case of the rotating equilibrium this term vanishes at the
endpoints since $\gamma_0(\pm 1) = 0$, provided $\zeta(\Gamma(x)) \in
C^1([-1,1])$. Under the same condition on $\zeta(\Gamma(x))$, the
inverse square-root singularity of the circulation density in
\eqref{eq:gamma0t} is preserved in the case of the translating
equilibrium. We thus conclude that the two constraints discussed above
are satisfied automatically by perturbations $\zeta(\Gamma(x))$ that
are continuously differentiable functions of $x$, such as polynomials.
Hence there are no extra constraints to add to the eigenvalue problems
\eqref{eq:evpBRr} and \eqref{eq:evpBRt}.

\subsection{Solution of Eigenvalue Problems \eqref{eq:evpBRr} and \eqref{eq:evpBRt}}
\label{sec:evp}

In this section we present closed-form solutions to eigenvalue
problems \eqref{eq:evpBRr} and \eqref{eq:evpBRt} corresponding to the
rotating and translating equilibria. We remark that the form of these
solutions was inspired by solutions to these problems obtained
numerically using a spectral Chebyshev method, with
  complementary insights provided by Galerkin and collocation
formulations \citep{b01}.

\subsubsection{Eigenvalue Problem for the Rotating Equilibrium}
\label{sec:evpr}

\begin{figure}
\centering 
\mbox{
\subfigure[]{\includegraphics[bb=0 0 408 284, width=0.4\textwidth]{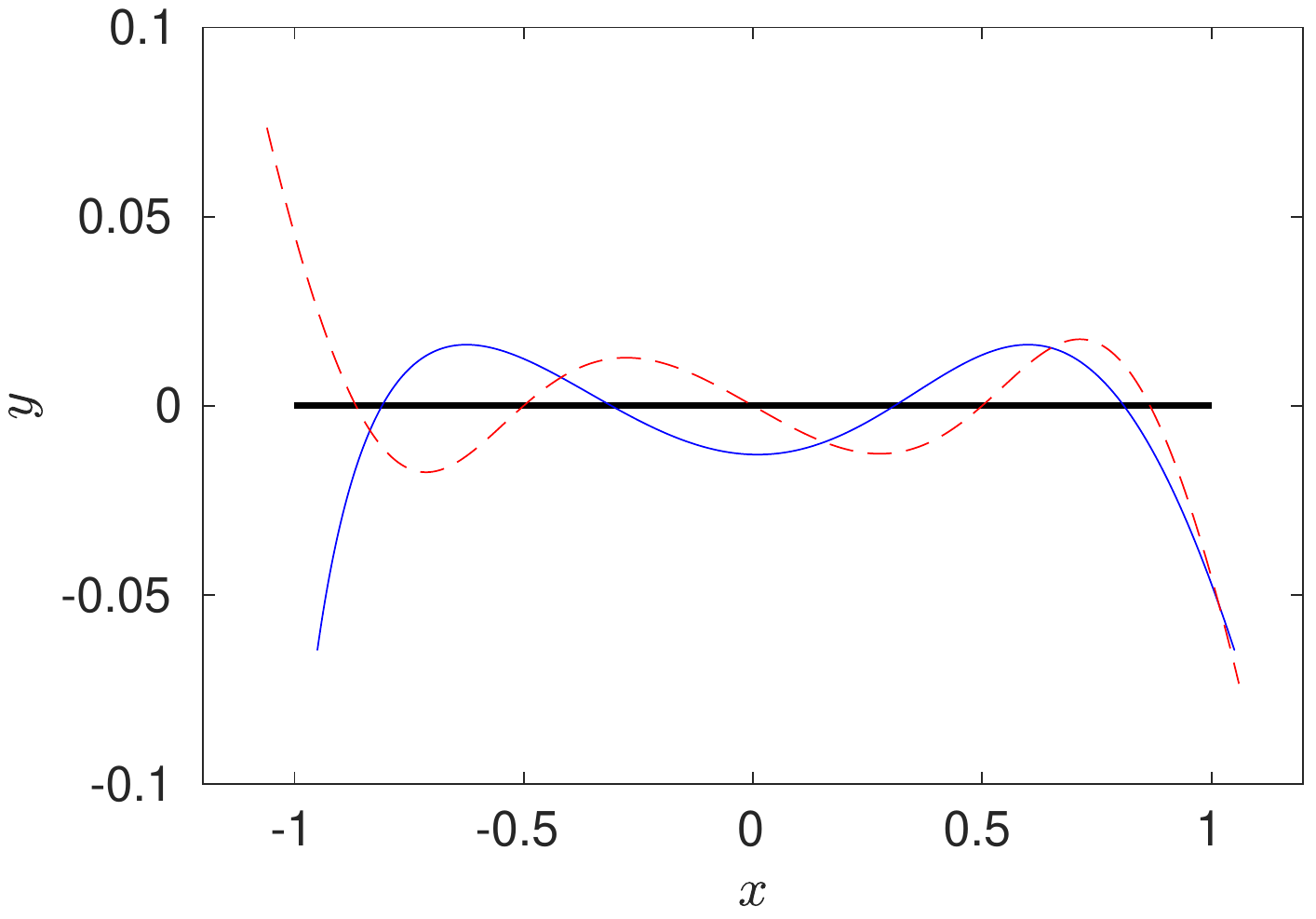}} \qquad
\subfigure[]{\includegraphics[bb=0 0 408 284, width=0.4\textwidth]{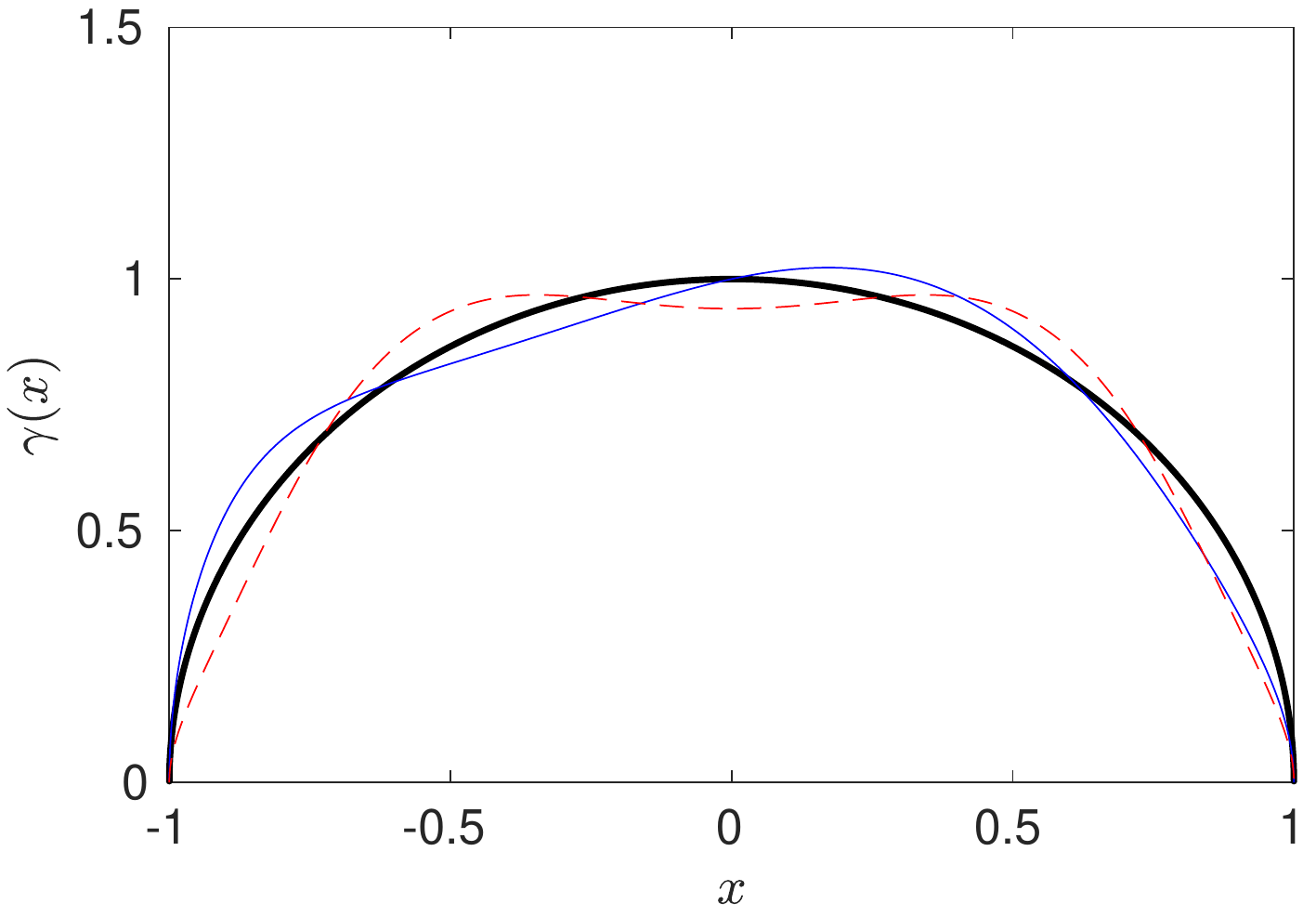}}}
\mbox{
\subfigure[]{\includegraphics[bb=0 0 408 284, width=0.4\textwidth]{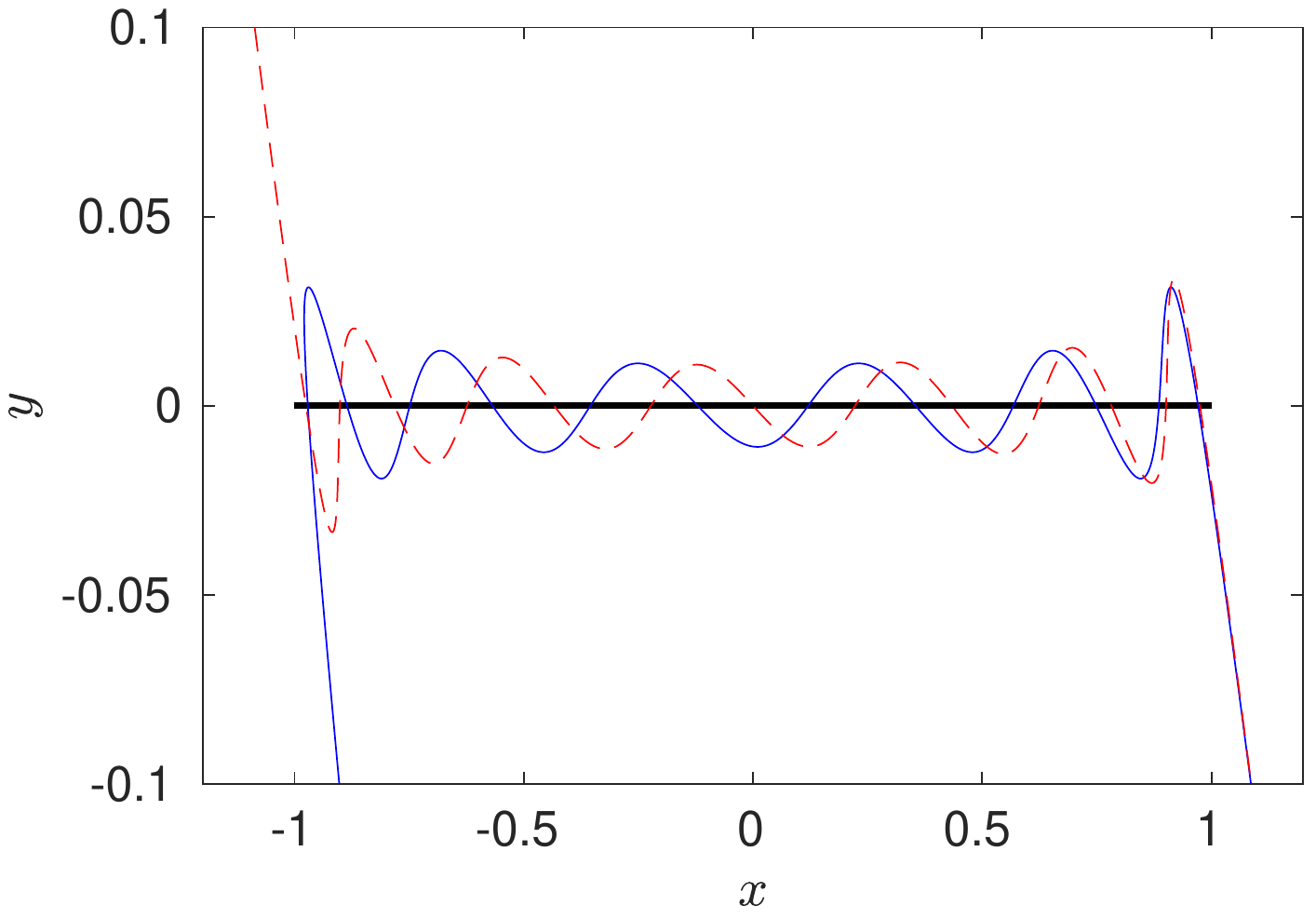}} \qquad
\subfigure[]{\includegraphics[bb=0 0 408 284, width=0.4\textwidth]{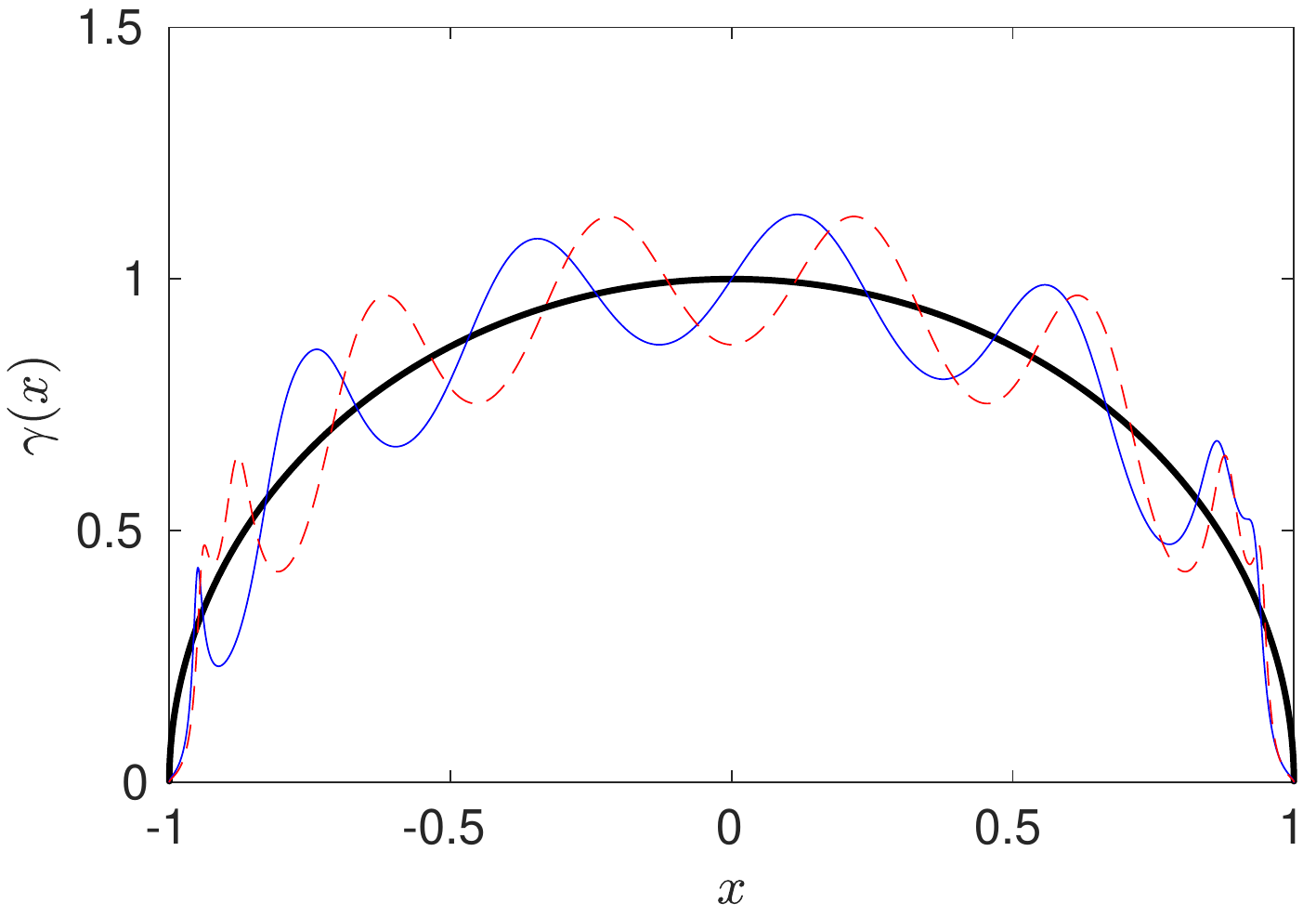}}}
\mbox{
\subfigure[]{\includegraphics[bb=0 0 408 284, width=0.4\textwidth]{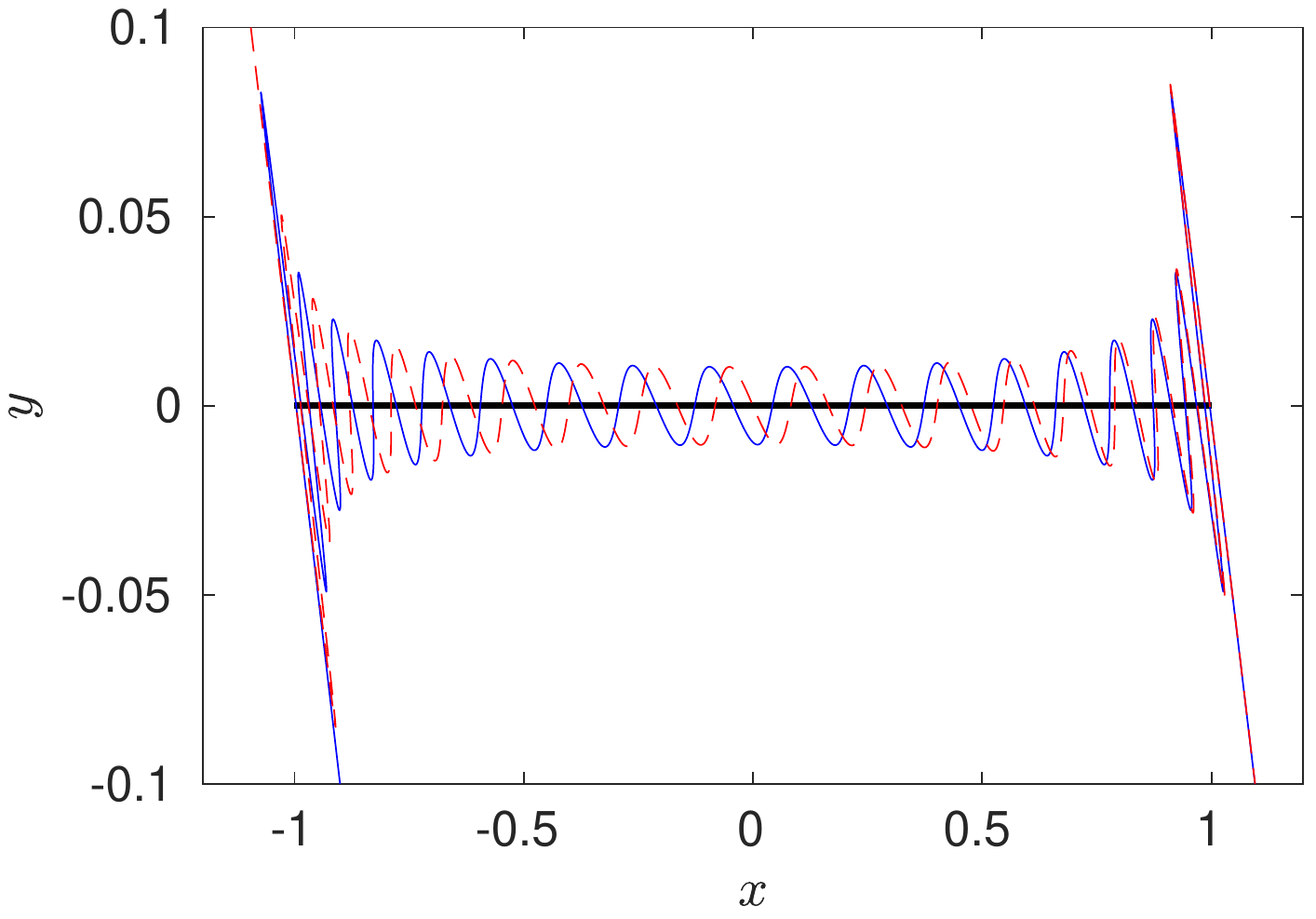}} \qquad
\subfigure[]{\includegraphics[bb=0 0 408 284, width=0.4\textwidth]{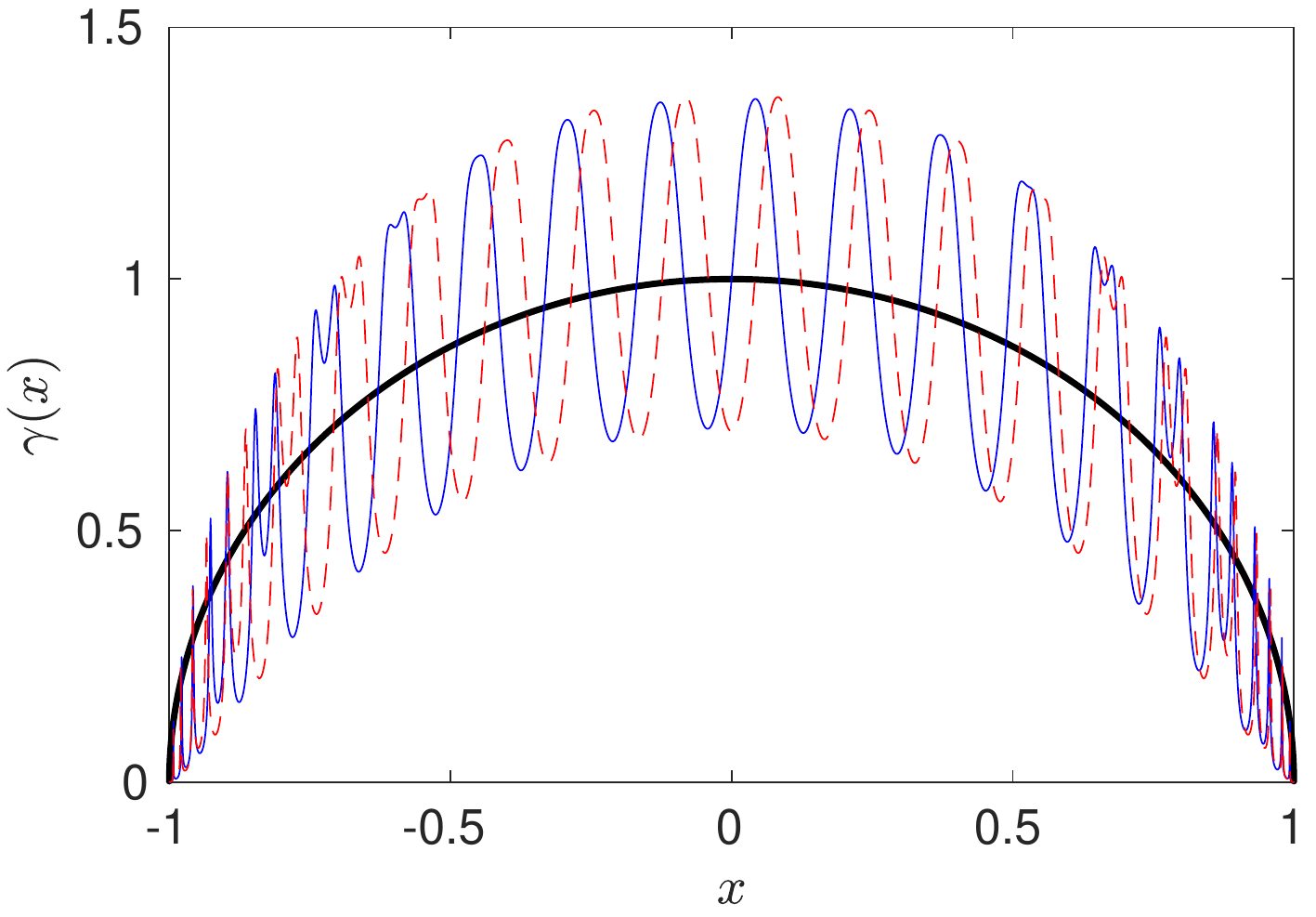}}}
\caption{Unstable eigenvectors of the rotating equilibrium
  corresponding to (a,b) $k=4,5$, (c,d) $k=12,13$ and
  (e,f) $k=36,37$ in expressions \eqref{eq:EVrc}.  The left
  column shows the perturbed sheet geometry $Z(\Gamma(x))$, and the
  right column the corresponding circulation density $\gamma(x)$, with
  $\epsilon = 10^{-2}$. Solid blue and dashed red lines represent,
  respectively, the perturbations corresponding to even and odd values
  of $k$, while thick black lines represent the equilibrium
  configuration.}
\label{fig:EVr}
\end{figure}

We begin by expressing the hypersingular operator defined in
\eqref{eq:H} in terms of integrals defined in Cauchy's principal-value
sense as follows:
\begin{align}
(\H \zeta)(\Gamma(x)) & = \frac{1}{2\pi} \mbox{f.p.} \int_0^{{\Gt}}
\frac{\zeta(\Gamma(x)) - \zeta(\Gamma')}{[x - x(\Gamma')]^2}\, \dd\Gamma'
= \frac{1}{2\pi} \mbox{f.p.} \int_{-1}^1
\frac{\zeta(\Gamma(x)) - \zeta(\Gamma(\xi))}{[x - \xi]^2} \gamma_0(\xi)\,\dd \xi \nonumber \\
&= - \frac{1}{2\pi} \left[ \zeta(\Gamma(x)) \frac{\dd}{\dd x} \pvint_{-1}^1
\frac{2\sqrt{1-\xi^2}}{x - \xi}\, \dd\xi - \frac{\dd}{\dd x} \mbox{p.v.}  \int_{-1}^1
\frac{\zeta(\Gamma(\xi)) 2\sqrt{1-\xi^2}}{x - \xi}\,\dd\xi \right], 
\nonumber \\
&= - \zeta(\Gamma(x)) + \frac{1}{\pi} \frac{\dd}{\dd x} \pvint_{-1}^1
\frac{\zeta(\Gamma(\xi)) \sqrt{1-\xi^2}}{x - \xi}\,\dd\xi,
\label{eq:Hzeta}
\end{align}
where we have used \eqref{eq:gamma0r} and the identity $\pvint_{-1}^1
\sqrt{1-\xi^2}(x - \xi)^{-1} \,\dd\xi = \pi x$. Next, applying this
operator to the Chebyshev polynomial of the second type $U_k$ yields
\begin{equation}
(\H U_k)(x) = - U_k(x) + \frac{\dd}{\dd x} T_{k+1}(x) = k U_k(x), \qquad k=0,1,\dots,
\label{eq:HUk}
\end{equation}
where $T_k$ is the Chebyshev polynomial of the first kind.  We have
also used the identities $\pvint_{-1}^1 \sqrt{1-\xi^2}U_{k-1}(\xi)(x -
\xi)^{-1}\,\dd\xi = \pi T_k(x)$ and $\dd T_{k}/\dd x = k U_{k-1}$
valid for $k \ge 1$ \citep{NIST:DLMF}. Relation \eqref{eq:HUk} implies
that $k=0,1,\dots$ and $U_k$ are, respectively, the eigenvalues and
eigenvectors of the operator $\H$ in \eqref{eq:H}.  We then rearrange
problem \eqref{eq:evpBRr} as
\begin{equation}
- \lambda^2 \hzeta^r = - (\I - \H)(\I + \H) \hzeta^r = (\H^2 - \I^2) \hzeta^r,
\label{eq:evpBRr2}
\end{equation}
where $\I$ denotes the identity operator. Evidently,
$\hzeta^r(\Gamma(x)) = U_k(x)$ is also an eigenfunction of problem
\eqref{eq:evpBRr2} with the eigenvalue $\lambda_k = \pm \ci \sqrt{k^2 -
  1}$. Since $\hzeta^i(\Gamma(x))$ satisfies an equation identical to
\eqref{eq:evpBRr2}, the solution of eigenvalue problem \eqref{eq:evpBRr}
is $\hzeta^r_k(\Gamma(x)) = U_k(x)$, $\hzeta^i_k(\Gamma(x)) = \theta_k U_k(x)$. Inserting this representation into \eqref{eq:evpBRr} leads to
$\theta_k = \sqrt{(k+1)/(k-1)}$ for $k=2,3,\dots$.
The cases with $k=0,1$ need to be considered separately: the
corresponding solutions can be easily deduced from \eqref{eq:evpBRr}.
Thus the eigenvalues $\lambda_k$ and eigenvectors $\hzeta_k =
\hzeta^r_k+ \ci\hzeta^i_k$ of the problem \eqref{eq:evpBRr} are
\begin{subequations}
\label{eq:EVr}
\begin{alignat}{2}
\lambda_0 &= \pm 1,& \qquad \hzeta_0(\Gamma) &= 1 \pm \ci^2 = 0, 2, \label{eq:EVra} \\
\lambda_1 &= 0,& \qquad \hzeta_1(\Gamma) &= \ci x(\Gamma), \label{eq:EVrb} \\
\lambda_k &= \pm \ci \sqrt{k^2 -  1},&  \hzeta_k(\Gamma) & = \left(1 \pm \ci \sqrt{\frac{k+1}{k-1}}\right) U_k(x(\Gamma)), \ k=2,3,\dots. \label{eq:EVrc} 
\end{alignat}
\end{subequations}
The neutrally-stable mode \eqref{eq:EVra} represents harmonic
oscillation of the centre of rotation around the origin. Mode
\eqref{eq:EVrb} associated with the zero eigenvalue represents the
stretching or compression of the sheet, and can be therefore interpreted
as connecting the rotating equilibrium defined by
\eqref{eq:BR0r}--\eqref{eq:gamma0r} with a nearby equilibrium.
Finally, there exists a countably infinite family of linearly stable and
unstable eigenmodes involving deformation of the sheet. In the limit of large $k$, the eigenvalues behave
as $\lambda_k \sim \pm \ci k$. The eigenfunctions \eqref{eq:EVrc}
corresponding to three different even and odd values of $k$ are shown
in Figure \ref{fig:EVr} in terms of perturbed shapes and perturbed
circulation densities of the vortex sheet.

\subsubsection{Eigenvalue Problem for the Translating Equilibrium}
\label{sec:evpt}
Analogously to \eqref{eq:Hzeta}, the action of the hypersingular
operator $\G$ on the perturbation $\zeta$ can be expressed using
\eqref{eq:gamma0t} as
\begin{align}
(\G \zeta)(\Gamma(x)) & = \frac{1}{2\pi} \mbox{f.p.}\int_{-1}^1
\frac{\zeta(\Gamma(x)) - \zeta(\Gamma(\xi))}{[x - \xi]^2} \gamma_0(\xi)\,\dd \xi = \frac{1}{2\pi} 
\mbox{f.p.}\int_{-1}^1 \frac{\zeta(\Gamma(x)) - \zeta(\Gamma(\xi)}{[x - \xi]^2} \frac{2\xi}{\sqrt{1-\xi^2}}\, \dd\xi \nonumber \\
&= - \frac{1}{\pi} \left[ \zeta(\Gamma(x)) \frac{\dd}{\dd x} \mbox{p.v.}\int_{-1}^1
\frac{\xi}{\sqrt{1-\xi^2}(x - \xi)}\, \dd\xi - \frac{\dd}{\dd x}  \mbox{p.v.}\int_{-1}^1
\frac{\zeta(\Gamma(\xi))\xi}{\sqrt{1-\xi^2}(x - \xi)}\,\dd\xi \right], 
\nonumber \\
&=\frac{1}{\pi} \frac{\dd}{\dd x}  \mbox{p.v.}\int_{-1}^1
\frac{\zeta(\Gamma(\xi))\xi}{\sqrt{1-\xi^2}(x - \xi)}\,\dd\xi,
\label{eq:Gzeta}
\end{align}
where we have used the identity $\pvint_{-1}^1 \xi (1-\xi^2)^{-1/2} (x -
  \xi)^{-1} \,\dd\xi= -\pi$.  Representing the perturbation as a function of
$x$ in terms of a Chebyshev series expansion with complex coefficients
\begin{equation}
\zeta(\Gamma(x)) = \sum_{k=0}^{\infty} (\alpha_k+i \beta_k) T_k(x), \qquad \alpha_k,\beta_k \in \RR, \qquad k=0,1,\dots,
\label{eq:zetacheb}
\end{equation}
we have $(\G \zeta)(x) = \sum_{k=0}^{\infty} (\alpha_k+\ci \beta_k) (\G T_k)(x),$ where
\begin{equation}
(\G T_k)(x) = \frac{1}{\pi} \frac{\dd}{\dd x}  \mbox{p.v.}\int_{-1}^1
\frac{\xi T_k(\xi)}{\sqrt{1-\xi^2}(x - \xi)}\, \,\dd\xi, \qquad k \ge 0. \label{eq:GTk}
\end{equation}
Using the identity $\pvint_{-1}^1 T_k(\xi)(1-\xi^2)^{-1/2}(x
- \xi)^{-1} \,\dd\xi = - \pi U_{k-1}(x)$, $k \ge 1$, and the
recurrence relations characterizing the Chebyshev polynomials of the
first and second kind \citep{NIST:DLMF}, we find
\begin{subequations}
\begin{align}
(\G T_0)(x) &= 0, \label{eq:GT0} \\ 
(\G T_k)(x) &= \frac{1}{1-x^2} \left[ k x T_k(x) - U_{k-1}(x) \right], \qquad k = 1,2,\dots.
\label{eq:GTkk}
\end{align}
\end{subequations}
We then define 
\begin{subequations}
\begin{align}
\w(t)  & := \left[ \alpha_0(t),\alpha_1(t),\dots, \beta_0(t), \beta_1(t),\dots \right]^T, \label{eq:w} \\ 
[\bG]_{jk} &:= \int_{-1}^1
\frac{T_j(\xi) (\G T_k)(\xi)}{\sqrt{1-\xi^2}}\, \dd\xi, \quad j,k = 0,1,\dots,
\label{eq:Gjk}
\end{align}
\end{subequations}
where the last expression represents the $j$th Chebyshev coefficient
of $(\G T_k)(x)$. Then the system \eqref{eq:dBRt2} can be rewritten
as an infinite-dimensional vector equation
\begin{equation}
\frac{d}{dt} \w = \begin{bmatrix} 
\phantom{-}\0 & - \bG \\
-\bG  & \phantom{-}\0 \end{bmatrix} \w =: \bA \w,
\label{eq:A}
\end{equation}
where $\0$ represents the null matrix. We remark that when the
operator $\G$ acts on a polynomial, the result is a polynomial of
degree reduced by one.  Thus, matrix $\bG$ representing this operator
in the Chebyshev basis is upper triangular with
zeros on its main diagonal. Therefore, we conclude that zero is
the only eigenvalue of the operator $\G$ and hence also of the eigenvalue
problem \eqref{eq:evpBRt}. The relation \eqref{eq:GT0} indicates that
$\hzeta^r(\Gamma) = \hzeta^i(\Gamma) = 1$ is the only eigenvector,
which implies that the eigenvalue $\lambda = 0$ has infinite algebraic
multiplicity and geometric multiplicity equal to 1. The matrix $\bA$
is nilpotent of degree infinity.

In the presence of such an extreme form of degeneracy, solutions of
system \eqref{eq:A} corresponding to some initial condition $\w_0 \in
l^2$ can be written as \citep{Perko2008} 
\begin{equation}
\w(t)  = \ee^{t \bA}\w_0 = \bP \diag\{\ee^{\lambda_j t} \} \bP^{-1} \left[
\bI + t \bA + \frac{t^2}{2} \bA^2  + \dots \right] \w_0
 =  \left[
\bI + t \bA + \frac{t^2}{2} \bA^2  + \dots \right] \w_0,
\label{eq:wt}
\end{equation}
where $\bI$ is the identity matrix and $\bP := [\v_0, \v_1, \dots ]$ is
a matrix with columns given by the generalized eigenvectors $\v_k$,
$k=0,1,\dots$ obtained from the Jordan chain \citep{Perko2008}
\begin{subequations}
\label{eq:Av}
\begin{align}
\bA \v_1  & = \v_0 = \left[ 1,0,\dots, 1,0,\dots \right]^T, \label{eq:Av1} \\ 
\bA \v_{k+1}  & = \v_k, \qquad k = 1,2,\dots. \label{eq:Avk}
\end{align}
\end{subequations}
Because of property \eqref{eq:GTkk}, the Jordan chain of generalized
eigenvectors consists of polynomials of degree increasing with $k$, and
$\v_k$ represents the Chebyshev coefficients of a polynomial of degree
$k$. Since the generalized eigenvectors are linearly independent, the
matrix $\bP$ is invertible. As all the eigenvalues are equal to zero,
the product of the first three factors on the first line in expression
\eqref{eq:wt} reduces to the identity matrix.  Expanding the initial
condition in terms of the generalized eigenvectors from the Jordan
chain \eqref{eq:Av} as $\w_0 = \eta_0 \v_0 + \eta_1 \v_1 + \dots$ for
some $\eta_0, \eta_1, \dots \in \RR$ and using the property
that $\bA^j \v_k = \0$ for $0 \le k < j$, we can rewrite the solution
\eqref{eq:wt} as
\begin{equation}
\w(t) = \w_0  + t (\eta_1 \v_1 + \eta_2 \v_2 + \dots )
              + \cdots
              + \frac{t^n}{n!} (\eta_n \v_n + \eta_{n+1} \v_{n+1} + \dots )
              + \cdots, \qquad 2 < n < \infty. 
\label{eq:wt2}
\end{equation}
This form of the solution allows us to conclude that, for each integer
$n>0$, there exists a perturbation $\w_0$ given by a polynomial of
degree equal to or greater than $n$ which grows in time at a rate
proportional to at least $t^n$.

\begin{figure}
\centering
\includegraphics[bb=0 0 408 284, width=0.6\textwidth]{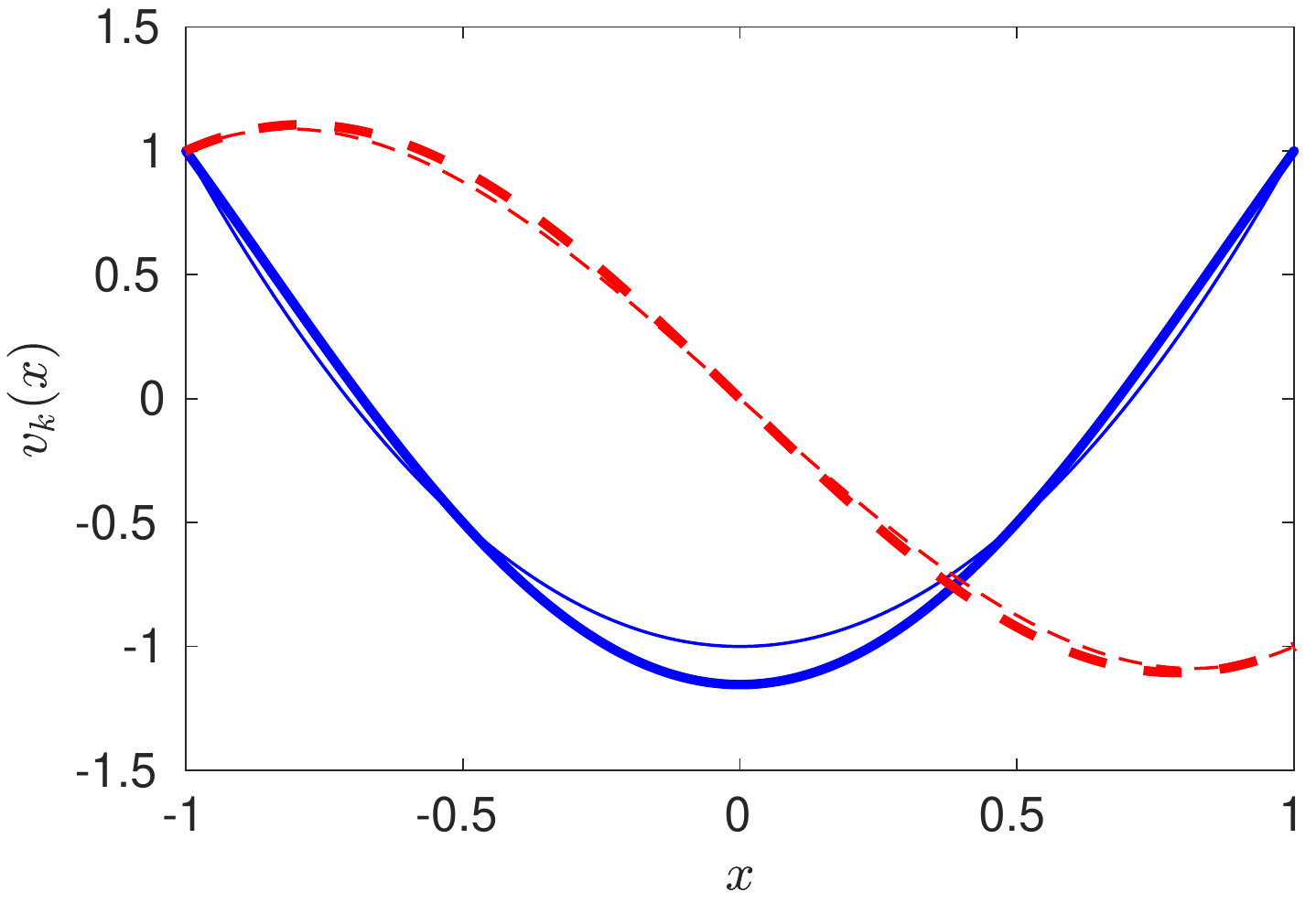}
\caption{Generalized eigenvectors $v_2$, $v_4$ (blue solid lines) and
  $v_3$, $v_5$ (red dashed lines) as functions of $x$. Thicker lines
  represent generalized eigenvectors of a higher degree. The
  graphs of the remaining generalized eigenvectors $v_6$
  ,$v_7$, $\dots$ are essentially indistinguishable from the thicker
  curves.}
\label{fig:vk}
\end{figure}

In order to understand the structure of the fastest-growing
perturbations represented by \eqref{eq:wt2}, it is instructive to
examine the generalized eigenvectors defined in \eqref{eq:Av} as
functions of $x \in [-1,1]$, i.e.~$v_k(x) = \sum_{j=0}^k [\v_k]_j
T_j(x)$, $k=0,1,\dots$. The first six generalized eigenvectors then
take the form
\begin{subequations}
\label{eq:vk}
\begin{alignat}{2}
v_0(x) &= T_0(x) = 1, & \qquad v_1(x) &= T_1(x) = x, \\
v_2(x) &= \frac{1}{4}T_2(x) = \frac{1}{2} x^2-\frac{1}{4}, & \qquad v_3(x) &= \frac{1}{24}T_3(x) - \frac{5}{24}T_1(x) = \frac{1}{6} x^3-\frac{1}{3}x, \\
v_4(x) &= \frac{1}{192}T_4(x) - \frac{7}{96}T_2(x) & \qquad v_5(x) &= \frac{1}{1920}T_5(x) - \frac{9}{640}T_3(x) + \frac{61}{960}T_1(x)  \nonumber \\
 & = \frac{1}{24}x^4 - \frac{3}{16}x^2 + \frac{5}{64},& & =  \frac{1}{120}x^5 - \frac{1}{15}x^3 + \frac{13}{120}x
\end{alignat}
\end{subequations}
and some of them are plotted in Figure \ref{fig:vk}. The remaining
generalized eigenvectors follow the same pattern.  We observe that the
generalized eigenvectors of even order consist of even-degree
polynomials only and vice versa. In all cases the magnitude of the
coefficients decreases with the degree of the term, so that the form
of the generalized eigenvectors is dominated by their lower-degree
terms. As a result, while the generalized eigenvectors \eqref{eq:vk}
are linearly independent \citep{Perko2008}, they form a strongly
non-normal set, as shown in Figure~\ref{fig:vk}. This implies that
when the initial perturbation $\zeta(0,\Gamma(x))$ in the form of a
generic degree-$n$ polynomial of $x$ is expanded in terms of the
generalized eigenvectors \eqref{eq:vk}, the expansion coefficients
$\eta_k$, $k=0,1,\dots,n$ generically increase in magnitudes with $k$.
This means that the fastest-growing components of such an initial
perturbation will be given by the generalized eigenvector $v_n$, and
will grow at a rate proportional to $t^n$. We also remark that the
even-degree generalized eigenvectors shown in Figure~\ref{fig:vk} have
some resemblance to the form of the most amplified perturbations
observed during the time evolution of a perturbed Prandtl-Munk vortex.
{More precisely, while the linear stability analysis cannot
  predict the roll-up of the sheet near its endpoints which is driven
  by nonlinear effects, it does appear to capture the change of the
  global shape of the sheet as in Figure 2 in \citet{Krasny1987} and
  Figure 4 in \citet{DevoriaMohseni2018}.}  However, given the form
\eqref{eq:wt2} of the solution of the linearized problem, it is
impossible to make this statement more quantitative, e.g.~by comparing
the growth rates.

\section{Time-dependent straight vortex sheets}
\label{sec:timedep}

The relative equilibrium involving Kirchhoff's rotating ellipse has
been generalized by including the effect of linear velocity fields.
\citet{ms71} found steady states involving ellipses in a uniform
straining field, while time-dependent solutions in the presence of a
simple shear were investigated by \citet{Kida1981}. In this section
{we describe analogous generalizations of the rotating equilibria
  of the vortex sheet described in \S\,\ref{sec:Rot_Eq}, whereas in
  \S\,\ref{sec:Klim} it is demonstrated that some of these solutions in
  fact coincide with the infinite-aspect-ratio limits of the
  Moore-Saffman and Kida vortex-patch solutions \citep{ms71,Kida1981}. An
  alternative derivation of these solutions will be presented in
  Appendix \ref{sec:ONeil}.}


\subsection{Governing equations}
\label{sec:ath}

We now derive from first principles the equation of motion for a
single vortex sheet in the presence of a linear external flow given by
$F(z)=Az+B{\overline{z}}$ with $A, B \in \mathbb{C}$.  Our focus
is on solutions in which the vortex sheet retains the form of a
straight segment, but with varying length and inclination angle to the
coordinate axes.  Since the external flow should be divergence-free,
we have $\Re\,[(\partial_x-i\partial_y)F(z)]=\Re\,[B]=0$. Hence,
without loss of generality, we can set the parameters of the external
flow as $A=r\ee^{-\ci\theta_0}$ and $B=\ci\Omega$, where $r>0$,
$\Omega \in \mathbb{R}$ and $\theta_0 \in (0,2\pi)$.  This form of the
external flow is a generalization of the earlier studies mentioned:
Kida's case~\citep{Kida1981} corresponds to $\theta_0=0$, while
$\Omega=0$ leads to Moore's and Saffman's case \citep{ms71}.  The
evolution of the vortex sheet represented by the curve $\L(t) \in \CC$
is then governed by the augmented Birkhoff-Rott equation
\eqref{eq:BRr}, which becomes
 \begin{equation}
\Dpartial{\overline{z}}{t} = \frac{1}{2\pi\ci} \mbox{p.v.}\int_\L \frac{\gamma(w)}{z-w} \vert \dd w \vert + r\ee^{-\ci\theta_0} z + \ci \Omega \overline{z}, \qquad z \in \L.
\label{eq:BRz}
\end{equation}
We now assume that the vortex sheet has the form of a line segment and
therefore can be parameterized as 
\begin{equation}
\L(t) \; \colon \; z(t,s)= a(t)s\ee^{\ci\theta(t)}, \qquad \mbox{where $-1 \leq s \leq 1$}.
\label{eq:Lt}
\end{equation} 
The positive-valued function $a(t)$ represents the half-length of the
vortex sheet, while the real-valued function $\theta(t)$ gives the
angle between $x$-axis and the sheet.  Although the support $\L(t)$ of
the circulation density changes in time, the total circulation $\Gt$
carried by the vortex sheet must be conserved in time. Hence, we take
the circulation density to have a form analogous to
\eqref{eq:gamma0r}, with
\begin{equation}
\gamma(z) = \frac{2}{a(t)}\sqrt{1-\frac{z^2}{a^2(t)\ee^{2\ci\theta(t)}}}, \qquad z \in \L(t).
\label{eq:gammat}
\end{equation}
Then the total circulation is independent of $a(t)$ and $\theta(t)$,
and
\begin{equation}
\Gt =  \int_{\L(t)} \gamma(z)\vert \dd z \vert = 
\int_{-1}^1  \frac{2}{a(t)}\sqrt{1-\frac{a^2(t)s^2\ee^{2\ci\theta(t)}}{a^2(t)\ee^{2\ci\theta(t)}}}a(t) \,\dd s = \int_{-1}^1 2 \sqrt{1-s^2} \,\dd s = \pi.
\end{equation}
Equations for $a(t)$ and $\theta(t)$ are obtained by substituting
\eqref{eq:Lt} and \eqref{eq:gammat} into
\eqref{eq:BRz}.  The singular integral in \eqref{eq:BRz} then becomes
\begin{align}
\mbox{p.v.}\int_\gamma\frac{\gamma(w)}{z-w}\vert dw \vert & = 
\mbox{p.v.}\int_{-1}^1  \frac{2\sqrt{1-(s')^2}}{a(t)s\ee^{\ci\theta(t)} - a(t)s'\ee^{\ci\theta(t)}}\, \dd s'  \\
& =\frac{\ee^{-\ci\theta(t)}}{a(t)} \mbox{p.v.}\int_{-1}^1  \frac{2 \sqrt{1-(s')^2}}{s - s'}\,\dd s' = \frac{2 \pi s}{a(t)}\ee^{-\ci\theta(t)},
\end{align}
where we have used the fact that the Hilbert transform of $\sqrt{1-s^2}$ is
$\pi s$. After performing some elementary algebraic operations we
obtain
\begin{equation}
\begin{bmatrix}
\dot{a}(t) \\ \dot{\theta}(t)
\end{bmatrix} 
= \begin{bmatrix}
a(t)r\cos(2\theta(t)-\theta_0) \\  
\frac{1}{a^2(t)} - r\sin (2\theta(t)-\theta_0) - \Omega
\end{bmatrix}
.
\label{eq:ath}
\end{equation}
As shown in Appendix \ref{sec:ONeil}, this system can also be derived
using an approach {proposed by \citet{ONeil2018b,ONeil2018a} to
  construct equilibrium solutions involving vortex sheets.}

The well-posedness of system \eqref{eq:ath} is easily established.
Rewriting the first equation in \eqref{eq:ath} as $a^{-1} \dot{a} =
r\cos (2\theta(t)-\theta_0)$, we immediately see that
\begin{equation}
a(t) = a(0)\exp\left[r \int_0^t \cos ( 2\theta(s)-\theta_0) \,\dd s \right].
\end{equation}
Since $-1 \leqq \cos (2\theta(t)-\theta_0) \leqq 1$ for all $t \in
\mathbb{R}$, we have
\begin{equation*}
 0<  a(0) \exp\left( -rt\right) \leqq a(t) \leqq a(0) \exp\left( rt\right)  <\infty.
\end{equation*}
This means the solutions of \eqref{eq:ath} cannot blow up in finite
time, but unbounded growth is possible in infinite time, in the sense
that $a(t) \rightarrow \infty$ as $t \to \pm\infty$ as we shall see
below.
 
\subsection{{Analysis of the fixed points of the system \eqref{eq:ath}}}
\label{sec:ath0}

Fixed points of the system \eqref{eq:ath} are obtained directly by solving
the equations $\dot{a}=\dot{\theta}=0$. From $\dot{a}=ar\cos
(2\theta-\theta_0)=0$ it immediately follows that $\theta =
\theta_n:=\frac{\theta_0}{2}+\frac{\pi}{4}+\frac{n \pi}{2}$ for $n \in
\mathbb{Z}$. Since $\sin(2\theta_n-\theta_0)=(-1)^n$, $\dot\theta=0$
is equivalent to $a^{-2} = r + \Omega$ when $\theta =
\theta_{2m}$ and to $a^{-2} = -r + \Omega$ when
$\theta = \theta_{2m+1}$, $m \in \ZZ$.  Hence, when $\Omega >r$, we have the
following two families of steady states
\begin{subequations}
\label{eq:ath12}
\begin{align}
(a_{2m}, \theta_{2m}) &= \left( \frac{1}{\sqrt{r+\Omega}}, \frac{\theta_0}{2}+\frac{\pi}{4}+m\pi \right), \label{eq:ath1} \\
(a_{2m+1}, \theta_{2m+1}) &= \left( \frac{1}{\sqrt{-r+\Omega}}, \frac{\theta_0}{2}+\frac{\pi}{4}+\left(m+\frac{1}{2}\right)\pi \right), \qquad m \in \ZZ. \label{eq:ath2}
\end{align}
\end{subequations}
On the other hand, when $r>\Omega>-r$, there is only one family of
steady states given by \eqref{eq:ath1} and there are no steady states
when $\Omega < -r$.  Since in the fixed frame of reference considered
here the relative equilibrium discussed in \S\,\ref{sec:Rot_Eq} has
the form of a periodic solution, in the limit $r,\Omega \rightarrow
0^+$ the fixed points \eqref{eq:ath12} disappear to infinity.

{We now analyze trajectories} near the fixed points
\eqref{eq:ath12}. {We emphasize that this is not a stability
  analysis of the equations motion as was carried out in
  \S\,\ref{sec:evpr}; instead, here we focus} on perturbations which
only affect $a(t)$ and $\theta(t)$ in \eqref{eq:Lt}, i.e.~those
that leave the vortex sheet in the form of a straight segment. The
Jacobian of system \eqref{eq:ath} is given by
\begin{equation}
\begin{bmatrix}
r \cos(2\theta-\theta_0) & -2ar\sin(2\theta-\theta_0) \\
-2 a^{-3} & -2r\cos(2\theta-\theta_0)
\end{bmatrix}.
\label{eq:Df}
\end{equation}
Computing the eigenvalues of Jacobian \eqref{eq:Df} evaluated at the critical
points yields
\begin{itemize}
\item $\lambda = \pm 2\sqrt{r(r+\Omega)}$ for the critical points
  \eqref{eq:ath1} when $r+\Omega>0$, indicating that these critical
  points are saddles,
\item $\lambda = \pm2\sqrt{r(-r+\Omega)} \, i$ for the critical points
  \eqref{eq:ath2} when $\Omega>r$, indicating that these critical
  points are centres.
\end{itemize}
The structure of the phase space $(a,\theta)$ of system \eqref{eq:ath}
for different combinations of the parameters $r$ and $\Omega$ is
explored in the next section.


\subsection{Phase plots}
\label{sec:phaseplots}

\begin{figure}
\begin{center}
\subfigure[]{\includegraphics[width=0.45\textwidth]{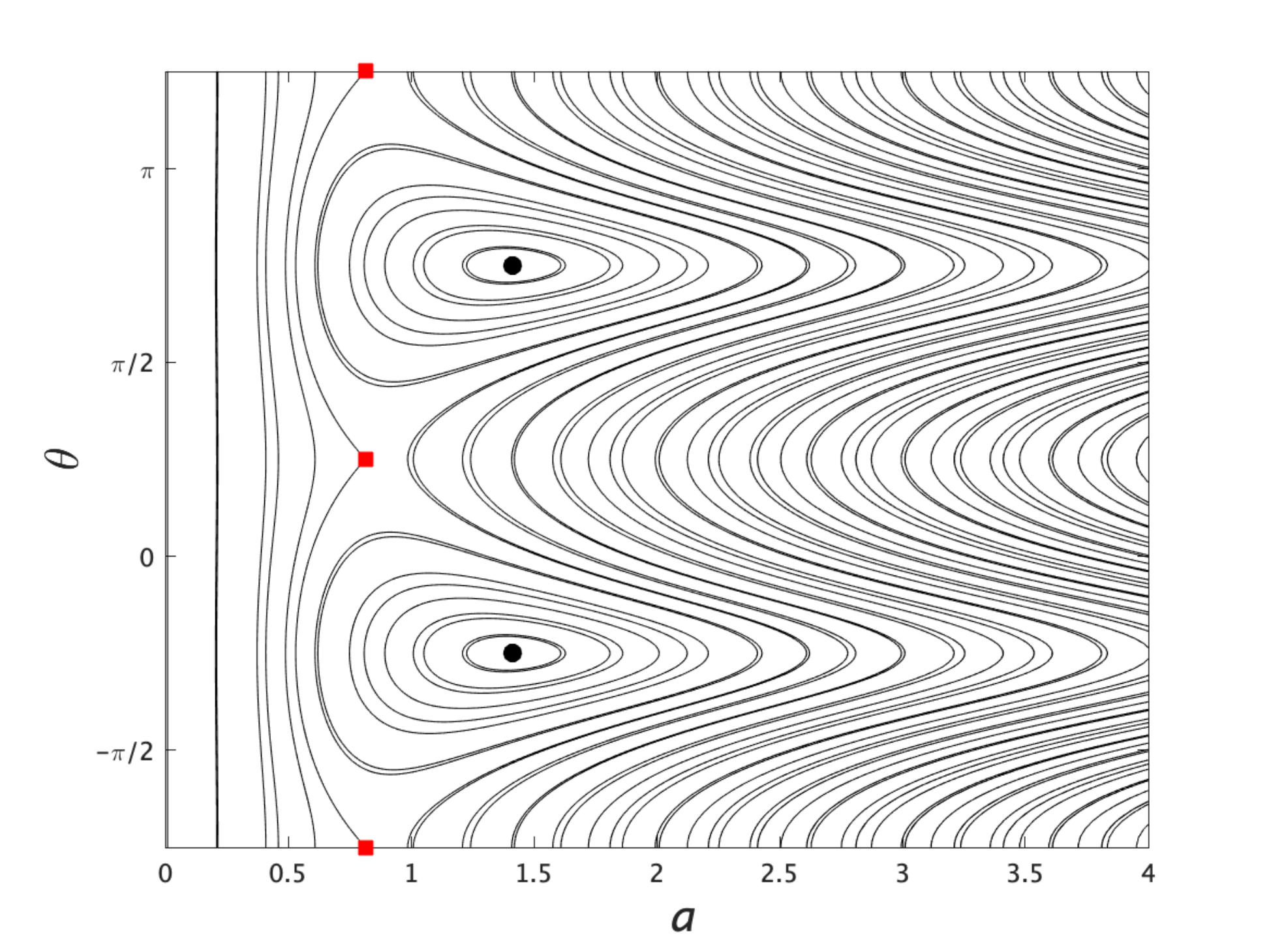}}
\subfigure[]{\includegraphics[width=0.45\textwidth]{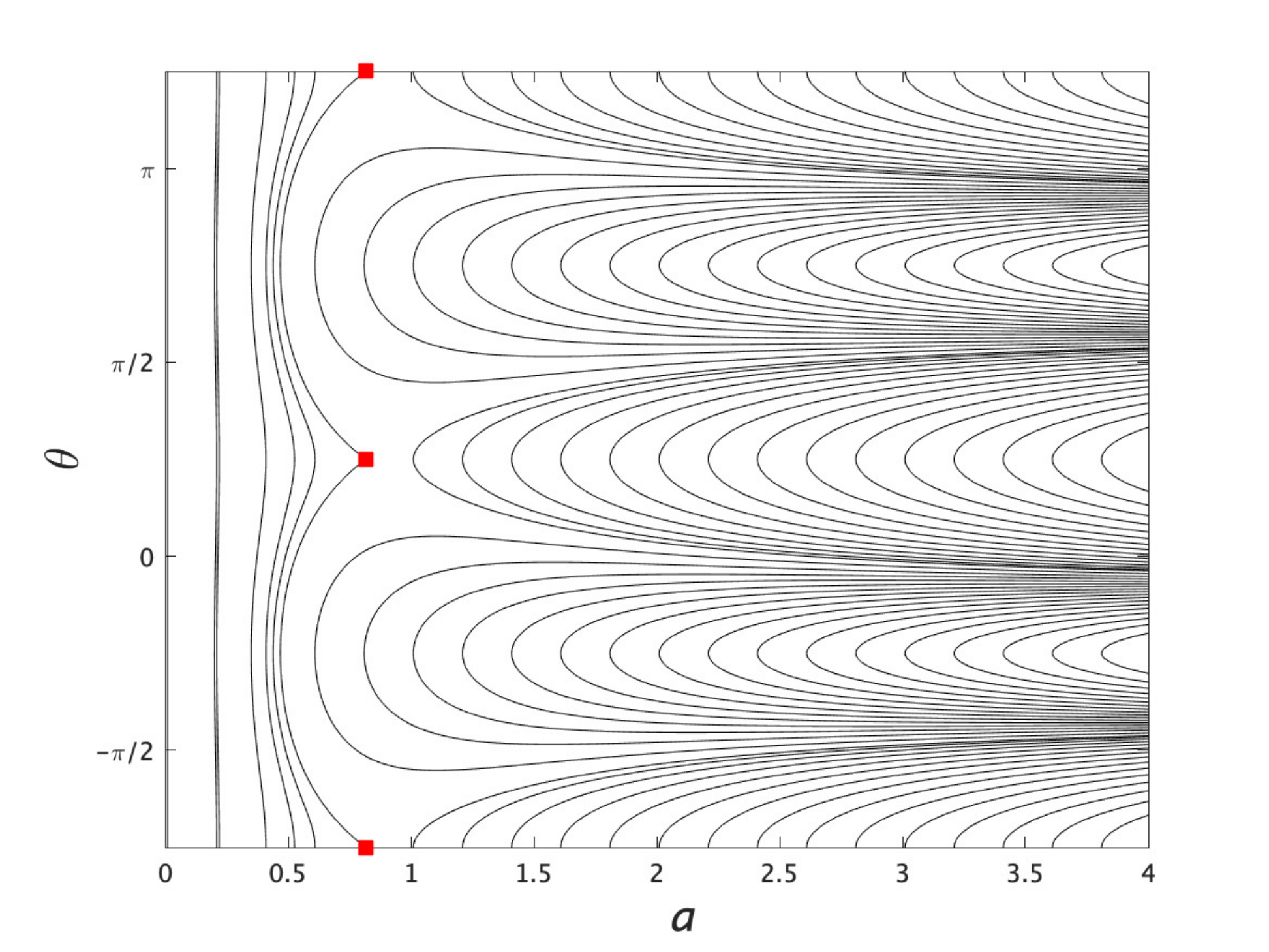}}
\subfigure[]{\includegraphics[width=0.45\textwidth]{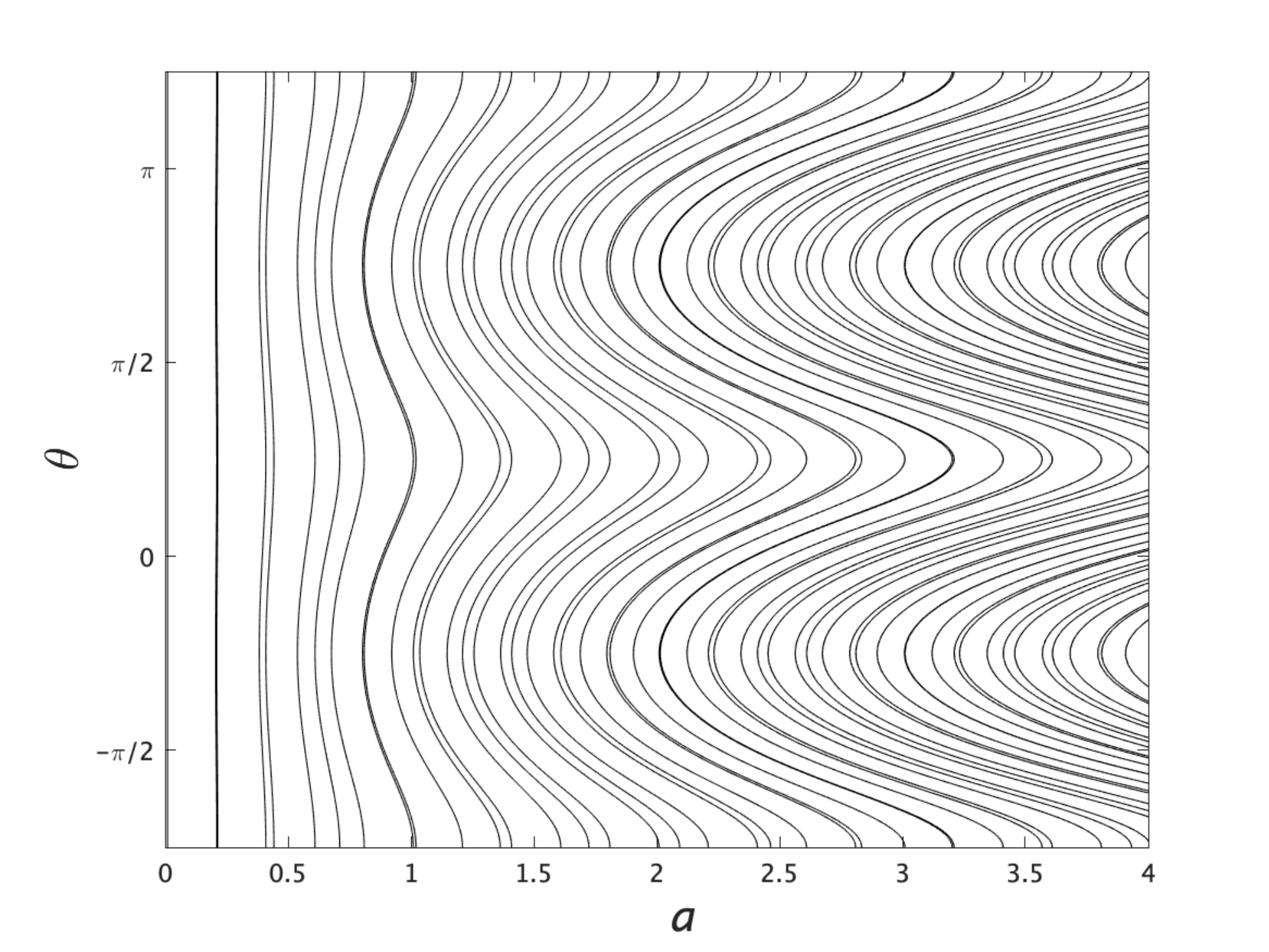}}
\end{center}
\caption{Solution trajectories of system \eqref{eq:ath} in the phase
  space $(a,\theta) \in \RR_+ \times \RR$ for different choices of
  parameters: (a) $(r,\Omega)=(0.5,1.0)$, (b) $(r,\Omega)=(1.0,0.5)$
  and (c) $(r,\Omega)=(0.5,-1.0)$. The black and red solid symbols
  represent centres and saddles as given in \eqref{eq:ath1} and
  \eqref{eq:ath2}. }
\label{fig:phase}
\end{figure}

The phase space $(a,\theta) \in \RR_+ \times \RR$ of the system
\eqref{eq:ath} is characterized by solving it numerically with
different initial conditions $(a(0),\theta(0))$. The results are shown
in Figure \ref{fig:phase} for the following three cases: (a)
$\Omega>r$, (b) $r>\Omega>-r$ and (c) $-r>\Omega$. Without loss of
generality, we choose $\theta_0=0$, since this parameter controls only
the inclination angle of the equilibrium configurations and not their
stability.

When $r=0.5$ and $\Omega=1.0$, the steady states $(a_{2m},
\theta_{2m})$ are centres and those at $(a_{2m+1}, \theta_{2m+1})$ are
saddles with heteroclinic connections; see Figure \ref{fig:phase}(a).
In the neighborhood of the centres the orbits are periodic
representing oscillation of the sheet without rotation. Outside the
heteroclinic connections, the solutions involve rotation of the sheet.
The direction of rotation for initial data located to the left of the
heteroclinic orbits is opposite to that for initial data located to
the right of the heteroclinic orbits.

For $r=1.0$ and $\Omega=0.5$, the steady states at $(a_{2m+1},
\theta_{2m+1})$ are saddle points linked by heteroclinic connections,
as seen in Figure \ref{fig:phase}(b). Orbits to the left of the
heteroclinic connections represent periodic solutions for which the
sheet rotates in the counter-clockwise direction while its length
oscillates. This is because in the second equation in system
\eqref{eq:ath} we obtain $\dot{\theta} \sim \frac{1}{2a^2}>0$ for
sufficiently small $a$. On the other hand, orbits to the right of the
heteroclinic connection represent unbounded solutions in which the
length of the vortex sheet goes to infinity as $t\rightarrow \infty$
while the inclination angle $\theta$ asymptotically approaches a
constant angle $\theta_\infty := -\frac{\pi}{12}+\frac{m\pi}{2}$, $m
\in \ZZ$, which satisfies the relation $\Omega+r\sin
2\theta_\infty=0$.

For $r=0.5$ and $\Omega=-1.0$, we observe only periodic orbits in
which the vortex sheet is rotating in the counter-clockwise direction;
see Figure \ref{fig:phase}(c). Longer sheets exhibit a more
significant variation of their length during one period of rotation.

We reiterate that in the analysis presented in this section we
restricted our attention to those solutions only where the sheet
retains the form of a straight segment with variable length and
inclination angle. Determining the effect of external fields on
motions of the vortex sheet involving arbitrary deformations remains
an open problem.

\section{Relation between Rotating Sheets and Ellipses}
\label{sec:relation}

\subsection{Stability Analysis Based on Limit of Kirchhoff's Rotating Ellipse}
\label{sec:Kstab}

\begin{figure}
\centering
\includegraphics[width=0.6\textwidth]{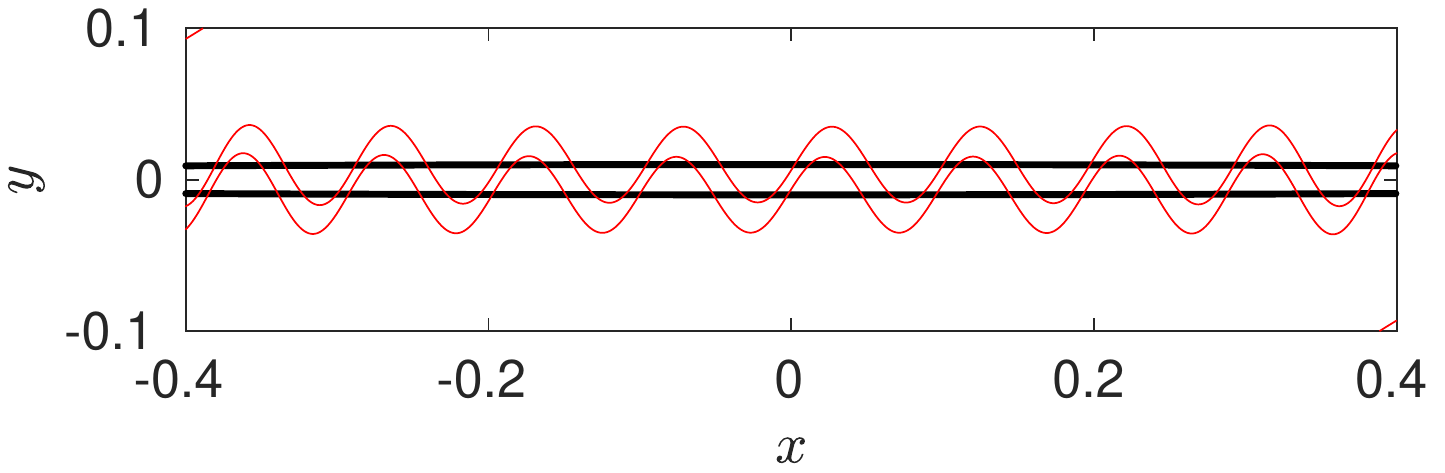}
\caption{(Thick solid line) a section of Kirchhoff's elliptic
    vortex with aspect ratio 100 together with (thin red line) its
    deformation by the unstable mode with wavenumber $m = 64$.}
\label{fig:love}
\end{figure}

We first list stability results for Kirchhoff's ellipse following
\cite{l93} who first found instability for $a/b > 3$, where $a$ and
$b$ are, respectively, the semi-major and semi-minor axis of the
ellipse, and then investigate the limit of large aspect ratio $a/b$.
Following the notation of \cite{mr08}, the dimensional frequency,
$\lambda_*$, of a mode-$m$ disturbance satisfies
\begin{equation}
\lambda_*^2 = \frac{\omega^2}{4} \left[ \left(\frac{2mab}{(a+b)^2} -
    1\right)^2  - \left(\frac{a-b}{a+b}\right)^{2m} \right],
\label{eq:RMl2}
\end{equation}
where $\omega$ is the value of the constant vorticity inside the
ellipse and $m > 0$. Now consider the limit of (\ref{eq:RMl2}) as
$b/a$ tends to $0$, with the circulation, $\Gt = \pi \omega ab$, kept
constant. This leads to
\begin{equation}
\lambda_*^2 = \frac{\omega^2b^2}{a^2} (2m-m^2) + o(1).
\end{equation}
This is negative for $m > 2$, so that modes with $m > 2$ are unstable.
The growth rate increases with $m$, a characteristic sign of
ill-posedness. We nondimensionalize $\lambda_*$ using the dimensional
angular velocity $\Omega_* = \omega ab/(a+b)^2$ of the ellipse. This
angular velocity tends to $\omega b/a$ as $b/a \to 0$, so we obtain the
nondimensional frequency
\begin{equation}
\lambda = \frac{\lambda_*}{\Omega_*} = \pm \ci \sqrt{m^2 - 2m} = \pm \ci [(m-1)^2 - 1]^{1/2}.
\end{equation}
To relate this result to the vortex sheet frequencies given by
\eqref{eq:EVra}--\eqref{eq:EVrc}, we note that the Cartesian mode number
$k$ is related to the azimuthal mode number $m$ by $m = k+1$, as in
(16a--b) of \cite{mr08}. We see that the unstable growth rates with $m
> 2$ correspond to (\ref{eq:EVrc}). The neutral mode $m = 2$
corresponds to (\ref{eq:EVrb}). The stable oscillations with $m = 1$
correspond to (\ref{eq:EVra}). This limiting process is
  illustrated in Figure \ref{fig:love} where we show Kirchhoff's
  elliptic vortex with aspect ratio 100 together with its deformation
  by the highest-wavenumber unstable mode predicted by Love's analysis
  \citep{l93,mr08}, which for the given aspect ratio corresponds to $m
  = 64$. In Figure \ref{fig:love} we note the emergence of a
  deformation pattern in the form of a slanted wave which is also
  evident in Figures \ref{fig:EVr}a,c,e.

\subsection{Limit of Kirchhoff's ellipse in the presence of external fields}
\label{sec:Klim}

The generalizations of the rotating equilibrium of a vortex sheet
described in \S\,\ref{sec:Rot_Eq} can be obtained by considering
suitable limits of the evolution of Kirchhoff's ellipse in the
presence of external fields.  Our discussion of the generalizations of
Kirchhoff's elliptical vortex follows \S\,9.3 in \citet{saffman-1992}.
Given a prescribed strain $U - \ci V = (\ci e(t) + g(t)) z$,
where $e(t)$ and $g(t)$ are real-valued functions of time, there
exist time-dependent patch solutions with constant vorticity $\omega$
in the form of rotating ellipses whose semi-major and semi-minor axes
$a(t)$ and $b(t)$ vary with time and the semi-major axis makes an
angle $\theta(t)$ with the $x$-axis, as described by equations (19)--(20) in
\S\,9.3 of \citet{saffman-1992}. Taking the limit $b \to 0$ with
constant circulation $\Gt = \pi \omega a(t) b(t)$ so that $\omega
\to \infty$, we obtain the equations governing the evolution of the
vortex sheet in the prescribed strain in terms of its half-length
$a(t)$ and inclination angle $\theta(t)$ in the form
\begin{equation}
\dot a = -e a \sin{2\theta} + g a \cos{2\theta}, \qquad
\dot\theta = \frac{\Gt}{\pi a^2} - e \cos{2\theta} - g\sin{2\theta}.
\label{eq:S1920}
\end{equation}
The circulation density of the sheet is then $2\Gt
\sqrt{1-s^2/a(t)^2}/(\pi a(t))$, where $s \in [0,a(t)]$ is distance
from the origin. It is clear that in the absence of the external
strain ($e = g = 0$) we recover the rotating equilibrium discussed in
\S\,\ref{sec:Rot_Eq}.  Moreover, the equations are equivalent to
(\ref{eq:ath}) when we consider the single vortex sheet (\ref{eq:Lt})
with circulation $\Gt=\pi$ in steady external strain,
i.e.~$e=-r\sin\theta_0$ and $g=r\cos\theta_0$, with $\Omega=0$.

The Moore--Saffman solutions are steady states involving ellipses in a
uniform straining field \citep{ms71}. The corresponding vortex sheet
equilibrium can be obtained without loss of generality by taking $g =
0$ in \eqref{eq:S1920}, which yields $\dot{a} = 0$, $\theta = 0$ along
with $e = \Gt/(\pi a^2)$.  \citet{saffman-1992} points out that if
patches do not satisfy the appropriate condition, ``the vortex is
pulled out into a long thin ellipse along the principal axis of
extension.'' Since the vortex sheet already has zero thickness, such
circumstances will result in unbounded growth of the sheet length
$a(t)$ accompanied by the vanishing of its circulation density. The
effect of solid-body rotation represented by an extra term of the form
$-\ci\Omega_0 \overline{z}$ was considered by \citet{Kida1981}. In
terms of the evolution of the vortex sheet the only difference is an
extra term $\Omega$ in the equation for $\dot\theta$ in
\eqref{eq:S1920}, and this recovers \eqref{eq:ath} for the case of
steady strain with rotation $\Omega$.

\section{Discussion and Conclusions}
\label{sec:final}

In this study we have established a number of new results concerning
the stability of the rotating and translating equilibria of open
finite vortex sheets.  Some of these findings complement analogous
results already known for unbounded, periodic and circular vortex
sheets. The main difference between these two types of equilibria, and
at the same time the source of several technical difficulties here, is
the presence of the endpoints.


The stability {analysis of} rotating equilibria shows similar
behavior to straight periodic sheets \citep{saffman-1992} and circular
sheets \citep{MichalkeTimme1967}. More specifically, there is a
countably infinite family of unstable modes with growth rates
increasing with the wavenumber $k$, as shown in \eqref{eq:EVrc}. Away
from the endpoints and in the limit of large wavenumbers the
corresponding unstable eigenmodes resemble the unstable eigenmodes of
a straight periodic sheet which have the form $(1-\ci) \sin(k \xi)$,
$\xi \in [0,2\pi]$. More precisely, near the centre of the sheet the
unstable eigenmodes have the form of slanted sine and cosine waves.
The reason for this analogy can be understood by examining the
structure of the eigenvalue problem \eqref{eq:evpBRr} and the
hypersingular integral operator \eqref{eq:Hzeta}. We see that when the
eigenvalues $\lambda$ have large magnitude, the terms due to the
background rotation in \eqref{eq:evpBRr} are dominated by the other
terms. Moreover, when the integral operator $\H$ acts on
high-wavenumber perturbations $\zeta(\Gamma(x))$, the circulation
density $\gamma_0(x)$ in \eqref{eq:gamma0r} can be locally
approximated by a constant for $x$ away from the endpoints.  Thus, in
this limit, the structure of the eigenvalue problem \eqref{eq:evpBRr}
becomes similar to the structure of the eigenvalue problem
characterizing the stability of straight periodic vortex sheets
\citep{saffman-1992}. Therefore, we can conclude that rotating finite
sheets are subject to the same Kelvin--Helmholtz instability as
straight sheets, which becomes more severe at higher wavenumbers and
rendering this problem similarly ill-posed.

On the other hand, the solution of the stability problem for the
translating vortex sheet in \S\,\ref{sec:evpt} is more nuanced since,
as a result of the degeneracy of the eigenvalue problem
\eqref{eq:evpBRt} with the hypersingular integral operator
\eqref{eq:Gzeta}, this equilibrium sustains unstable modes growing at
an algebraic rather than exponential rate. However, this algebraic
growth rate can be arbitrarily large provided the perturbations vary
sufficiently rapidly in space. Thus, this problem is ill-posed in a
similar way to vortex sheets exhibiting the classical Kelvin-Helmholtz
instability.  As suggested by the form of the generalized eigenvectors
shown in Figure \ref{fig:vk}, this analysis captures the general form
of the instability actually observed in numerical computations
\citep{Krasny1987,DevoriaMohseni2018}, although direct comparisons are
made difficult by the fact that the computations relied on various
regularized forms of the Birkhoff-Rott equation \eqref{eq:BR}, while
no such regularization was used in our stability analysis.  The
Prandtl-Munk vortex is thus the only known equilibrium involving a
vortex sheet which does not have exponentially unstable modes.  This
property can be attributed to the fact that the corresponding
circulation density \eqref{eq:gamma0t} is not sign-definite, so that
the self-induced straining field exerts a stabilizing effect.

The results reported in \S\,\ref{sec:timedep} show that the rotating
equilibrium discussed in \S\,\ref{sec:Rot_Eq} is ``robust'' in the
sense that configurations involving straight sheets but with
time-dependent length and inclination angle also arise as solutions in
the presence of external fields. However, we remark that the results
presented in \S\,\ref{sec:ath0} do not represent a complete stability
analysis since they do not account for perturbations affecting the
shape of the sheet. Generalizing this analysis to account for such
shape-deforming perturbations is thus an open problem. For the
periodic solutions of Figure \ref{fig:phase}a, this could be done by
combining the methods from \S\,\ref{sec:stability} with Floquet
theory. Another interesting open question is whether the translating
equilibrium admits generalizations analogous to those discussed in
\S\,\ref{sec:timedep}.

The analysis presented in \S\,\ref{sec:relation} demonstrates that the
relation between the rotating vortex sheet and Kirchhoff's ellipse
stipulated by \citet{batchelor-1988} does not merely concern the form
of the equilibrium configurations, but also applies to their stability
properties. {That this should be the case seems nontrivial because in
  the infinite-aspect-ratio limit the form of the Euler equation used
  to describe vortex patches and its linearization lose validity.  The
  practical value of the stability results obtained in
  \S\,\ref{sec:stability} is their simple and explicit form making
  comparisons with stability analyses of other configurations such a
  straight infinite vortex sheet straightforward. In contrast, the
  expressions describing unstable modes of Kirchhoff ellipse obtained
  by \citet{l93} are rather complicated.}  The relation between the
evolution of unbounded sheets of finite and zero thickness was
considered by \citet{BakerShelley1990,BenedettoPulvirenti1992}.  An
interesting open question is whether there exists a family of
vortex-patch equilibria that will converge to the Prandtl-Munk vortex
in a certain limit. {Another open problem is to understand
  whether the equilibria considered here are unique in the class of
  configurations involving a single open finite vortex sheet.}

\section*{Acknowledgments}

The authors thank Kevin O'Neil for interesting discussions about his
approach and {anonymous reviewers for insightful and constructive
  comments on the paper}. The first author acknowledges partial
support through an NSERC (Canada) Discovery Grant. The third author
was partially supported by the JSPS Kakenhi (B) (\#18H01136), the
RIKEN iTHEMS program in Japan, and a grant from the Simons Foundation
in the USA.  The authors would also like to thank the Isaac Newton
Institute for Mathematical Sciences for support and hospitality during
the programme ``Complex analysis: techniques, applications and
computations'' where this work was initiated. This programme was
supported by EPSRC grant number EP/R014604/1.


\appendix 

\section{Derivation of solutions from \S\,\ref{sec:timedep} following
  O'Neil's formulation}
\label{sec:ONeil}

In this appendix we show that the solutions obtained in
\S\,\ref{sec:timedep} as generalizations of the rotating equilibrium
from \S\,\ref{sec:Rot_Eq} can be obtained in an entirely different
manner using the method of \citet{ONeil2018b,ONeil2018a}.

We begin with vortex sheet equilibria in the presence of uniform
strain without rotation. The velocity field due to the sheet $\L$ and the strain
field is
\begin{equation}
f(z) = \frac{1}{2\pi\ci} \int_\L \frac{\gamma(w)\overline{\tau}(w)}{z-w}
\,\dd w + r \ee^{-\ci \theta_0} z, \qquad z \notin \L.
\end{equation}
The argument of \citet{ONeil2018b,ONeil2018a} shows that the extension
of $f^2(z)$ in the finite plane including the sheet $\L$ is entire. At
infinity, we find
\begin{align}
f^2(z) &= \left[ r\ee^{-\ci \theta_0} z + \frac{1}{2\pi\ci z} \int_\L \gamma(w) \,\vert \dd w \vert +
  \O(\vert z \vert^{-2}) \right]^2 = r^2\ee^{-2\ci \theta_0} z^2 + \frac{\Gt r\mbox{e}^{-\ci \theta_0}}{\pi \ci} + \O(\vert z \vert^{-1}),  \nonumber \\
  &= r^2\ee^{-2\ci \theta_0} z^2 -\ci r \ee^{- \ci \theta_0} + \O(\vert z \vert^{-1}), \qquad \vert z \vert \rightarrow \infty, \label{eq:f2}
\end{align}
where $\Gt=\pi$ is the circulation along the sheet. By Liouville's
theorem, $f^2(z)$ is in fact equal to the sum of the constant term and
the term unbounded at infinity, i.e.~one drops the terms $\O(\vert z
\vert^{-1})$. In a steady state the endpoints of the sheet $\L$ must
be stagnation points. Parameterizing the points on the vortex sheet as
$w = a s \ee^{\ci\theta}$ with $-1 \leq s \leq 1$, we obtain at the
endpoints
\begin{equation}
f^2(\pm a \ee^{\ci\theta}) = r^2 a^2 \ee^{2\ci(\theta-\theta_0)} -\ci r \ee^{-\ci \theta_0} = 0,
\end{equation}
which gives rise to the steady states (\ref{eq:ath12}) with $\Omega=0$.

For the general non-stationary case, with the points of the vortex
sheet given by $z = a(t) s \ee^{\ci\theta(t)}$, $-1 \leq s \leq 1$, we
have
\begin{equation}
f(z) = \frac{1}{2\pi\ci} \int_\L \frac{\gamma(w)\overline{\tau}(w)}{z-w}
\,\dd w + r\ee^{-\ci \theta_0} z = [\ci\dot\theta + u(s)] \overline{z}.
\end{equation}
The term proportional to $\dot\theta$ represents rotation and is
expected. The real function $u(s)$ in the last term corresponds to the
tangential velocity along the contour resulting from its extension or
contraction. It needs to be obtained as part of the solution, but we
only need to satisfy the kinematic condition $u(1) = \dot a$ at the
endpoints of the sheet where $s = 1$.  Using the identity
$\overline{z} =\ee^{-2\ci\theta} z$ valid for $z \neq 0$ and employing
the same process as in \eqref{eq:f2} above shows that the function
$[f(z) - (\ci\dot\theta + u(s)) \ee^{-2\ci\theta} z]^2$ is
meromorphic. The limit $|z| \to \infty$ then gives
\begin{equation}
[f(z)  - (\ci\dot\theta + u) z]^2 = A^2 z^2 + \frac{\Gt A}{\ci \pi} = A^2z^2 - \ci A = 0,
\end{equation}
where $A = r\ee^{-\ci \theta_0} - (\ci \dot\theta + u(s))\ee^{-2\ci\theta}$ and $\Gt=\pi$. We now use the kinematic relation $u(1) = \dot a$
at $z = a \ee^{\ci\theta}$ and obtain
\begin{equation}
Aa^2\ee^{2\ci \theta} = \ci.
\end{equation}
Separating this relation into real and imaginary parts gives the
equations (\ref{eq:ath}) for $\dot a$ and $\dot\theta$ with $\Omega =
0$. We can also obtain an expression for $u(s)$ from $f(z)$ if
desired.



\end{document}